\PassOptionsToPackage{unicode}{hyperref}
\PassOptionsToPackage{hyphens}{url}
\documentclass[
]{article}
\usepackage{xcolor}
\usepackage[margin=1in]{geometry}
\usepackage{amsmath,amssymb}
\usepackage{iftex}
\ifPDFTeX
  \usepackage[T1]{fontenc}
  \usepackage[utf8]{inputenc}
  \usepackage{textcomp} 
\else 
  \usepackage{unicode-math} 
  \defaultfontfeatures{Scale=MatchLowercase}
  \defaultfontfeatures[\rmfamily]{Ligatures=TeX,Scale=1}
\fi
\usepackage{lmodern}
\ifPDFTeX\else
\fi
\IfFileExists{upquote.sty}{\usepackage{upquote}}{}
\IfFileExists{microtype.sty}{
  \usepackage[]{microtype}
  \UseMicrotypeSet[protrusion]{basicmath} 
}{}
\makeatletter
\@ifundefined{KOMAClassName}{
  \IfFileExists{parskip.sty}{%
    \usepackage{parskip}
  }{
    \setlength{\parindent}{0pt}
    \setlength{\parskip}{6pt plus 2pt minus 1pt}}
}{
  \KOMAoptions{parskip=half}}
\makeatother
\usepackage{color}
\usepackage{fancyvrb}

\DefineVerbatimEnvironment{Highlighting}{Verbatim}{commandchars=\\\{\}}
\usepackage{framed}
\definecolor{shadecolor}{RGB}{248,248,248}
\newenvironment{Shaded}{\begin{snugshade}}{\end{snugshade}}

\newcommand{\AttributeTok}[1]{\textcolor[rgb]{0.13,0.29,0.53}{#1}}

\newcommand{\CommentTok}[1]{\textcolor[rgb]{0.56,0.35,0.01}{\textit{#1}}}

\newcommand{\ControlFlowTok}[1]{\textcolor[rgb]{0.13,0.29,0.53}{\textbf{#1}}}

\newcommand{\DecValTok}[1]{\textcolor[rgb]{0.00,0.00,0.81}{#1}}

\newcommand{\FloatTok}[1]{\textcolor[rgb]{0.00,0.00,0.81}{#1}}
\newcommand{\FunctionTok}[1]{\textcolor[rgb]{0.13,0.29,0.53}{\textbf{#1}}}

\newcommand{\NormalTok}[1]{#1}

\newcommand{\OtherTok}[1]{\textcolor[rgb]{0.56,0.35,0.01}{#1}}

\newcommand{\SpecialCharTok}[1]{\textcolor[rgb]{0.81,0.36,0.00}{\textbf{#1}}}

\newcommand{\StringTok}[1]{\textcolor[rgb]{0.31,0.60,0.02}{#1}}

\usepackage{graphicx}
\makeatletter
\newsavebox\pandoc@box
\newcommand*\pandocbounded[1]{
  \sbox\pandoc@box{#1}%
  \Gscale@div\@tempa{\textheight}{\dimexpr\ht\pandoc@box+\dp\pandoc@box\relax}%
  \Gscale@div\@tempb{\linewidth}{\wd\pandoc@box}%
  \ifdim\@tempb\p@<\@tempa\p@\let\@tempa\@tempb\fi
  \ifdim\@tempa\p@<\p@\scalebox{\@tempa}{\usebox\pandoc@box}%
  \else\usebox{\pandoc@box}%
  \fi%
}
\def\fps@figure{htbp}
\makeatother
\NewDocumentCommand\citeproctext{}{}

\makeatletter
 \let\@cite@ofmt\@firstofone
 \def\@biblabel#1{}
 \def\@cite#1#2{{#1\if@tempswa , #2\fi}}
\makeatother
\newlength{\cslhangindent}
\newlength{\csllabelwidth}
\newenvironment{CSLReferences}[2] 
 {\begin{list}{}{%
  \setlength{\itemindent}{0pt}
  \setlength{\leftmargin}{0pt}
  \setlength{\parsep}{0pt}
  \ifodd #1
   \setlength{\leftmargin}{\cslhangindent}
   \setlength{\itemindent}{-1\cslhangindent}
  \fi
  \setlength{\itemsep}{#2\baselineskip}}}
 {\end{list}}
\usepackage{calc}

\newcommand{\CSLLeftMargin}[1]{\parbox[t]{\csllabelwidth}{\strut#1\strut}}
\newcommand{\CSLRightInline}[1]{\parbox[t]{\linewidth - \csllabelwidth}{\strut#1\strut}}

\providecommand{\tightlist}{%
  \setlength{\itemsep}{0pt}\setlength{\parskip}{0pt}}
\usepackage{threeparttable}

\usepackage{fancyhdr}
\usepackage{longtable}
\usepackage[export]{adjustbox}
\usepackage{subcaption}
\usepackage[margin=1in]{geometry}
\usepackage{amsmath}
\usepackage{array}

\usepackage{newunicodechar}
\newunicodechar{ε}{\ensuremath{\varepsilon}}

\usepackage{pdflscape}
\newcommand{\blandscape}{\begin{landscape}}
\newcommand{\elandscape}{\end{landscape}}

\usepackage{comment}

\usepackage{arydshln}

\usepackage{bookmark}
\IfFileExists{xurl.sty}{\usepackage{xurl}}{} 
\hypersetup{
  pdftitle={Transportability of Prognostic Markers: Rethinking Common Practices through a Sufficient-Component-Cause Perspective},
  pdfauthor={Mohsen Sadatsafavi, Gavin Pereira, Wenjia Chen},
  hidelinks,
  pdfcreator={LaTeX via pandoc}}

\title{Transportability of Prognostic Markers: Rethinking Common
Practices through a Sufficient-Component-Cause Perspective}
\author{Mohsen Sadatsafavi, Gavin Pereira, Wenjia Chen}
\date{}

\begin{document}
\maketitle
\begin{abstract}
Transportability, the ability to maintain performance across
populations, is a desirable property of markers of clinical outcomes.
However, empirical findings indicate that markers often exhibit varying
performances across populations. For prognostic markers that are
advertised as predictive risk equations for an outcome of interest,
oftentimes a form of updating is required when the equation is
transported to populations with different outcome prevalences. Here, we
revisit transportability of prognostic markers through the lens of the
foundational framework of sufficient component causes (SCC). We argue
that transporting a marker ``as is'' implicitly assumes predictive
values are transportable, whereas conventional prevalence adjustment
shifts the locus of transportability to accuracy metrics (sensitivity
and specificity). Using a minimalist SCC framework that decomposes risk
prediction into broad causal constituents, we show that both approaches
rely on strong assumptions about the stability of cause distributions.
An SCC framework instead invites making transparent assumptions about
how different causes vary across populations, leading to different
transportation methods. For example, in the absence of any external
information other than outcome prevalence, an impartial perspective can
assume all causes are responsible for change in prevalence, leading to a
new form of marker transportation. Numerical experiments demonstrate
that different transportability assumptions lead to varying degrees of
information loss, depending on the distribution of causes across
populations. An SCC perspective challenges common assumptions and
practices for marker transportability, and results in novel
transportability methods based on explicit assumptions on how different
causes vary across populations.
\end{abstract}

\pagestyle{plain}


\let\thefootnote\relax\footnotetext{From Faculty of Pharmaceutical Sciences and Faculty of Medicine, Vancouver, Canada (MS); School of Population Health, WHO Collaborating Centre for Climate Change and Health Impact Assessment, and enAble Institute, Faculty of Health Sciences, Curtin University, Perth, Australia (GP); Faculty of Medicine, Universitas Negeri Malang, Indonesia (GP); Saw Swee Hock School of Public Health, National University of Singapore, Singapore (WC)}

\let\thefootnote\relax\footnotetext{* Correspondence to Mohsen Sadatsafavi, Room 4110, Faculty of Pharmaceutical Sciences, 2405 Wesbrook Mall, Vancouver, British Columbia, V6T1Z3, Canada; email: mohsen.sadatsafavi@ubc.ca}

\clearpage

\newpage

\subsection{Background}\label{background}

The underlying premise of reporting on the performance of biomarkers,
tests, and prediction models (which we generally refer to as
``markers'') is that performance metrics are transportable from one
population to another. However, in practice, this premise more often
than not fails to hold. Research has repeatedly shown the performance of
markers can vary significantly from one population to another,
especially when the prevalence of the outcome of interest varies across
populations.\textsuperscript{1}--\textsuperscript{4} For example, risk
prediction models for cardiovascular diseases have shown substantial
degradation in performance when transported to a new population with
differing cardiovascular disease prevalence, leading to substantial risk
of harm (defined as the net benefit of using the model being lower than
not using it).\textsuperscript{5} The likelihood of harm was reduced
when models were updated to account for the difference in outcome
prevalences between their source and the target populations. Similar
findings have been reported for predicting the risk of exacerbations of
obstructive lung disease.\textsuperscript{6} In another example, the
performance of machine learning models for detecting pneumonia on chest
X-rays substantially declined in data from settings not used to train
the model.\textsuperscript{7}

When transporting a marker for an outcome to a new population, some
information about the characteristics of that population -- generally
referred to as case-mix -- is often available. One of the most common
pieces of information is the outcome prevalence. For example, cancer
registries often provide a good estimate of cancer risk in a population,
or population-based studies might offer robust estimates of the
prevalence of Alzheimer's disease. This poses the question as to how
such information can be used to revise our assessment of marker
performance in the new population. For example, if we assume the marker
will retain its positive and negative predictive values between the two
populations, information on outcome prevalence is effectively unused. On
the other hand, if we assume the marker will retain its sensitivity and
specificity between the two populations, then the information on
prevalence can be combined with sensitivity and specificity to calculate
predictive values in the target population. These assumptions correspond
to fundamentally different transportability methods.

For risk prediction models that return a quantitative estimate of
outcome risk given patient characteristics, some form of model revision
is often needed to correct for the under-estimation or over-estimation
of risks in the new population.\textsuperscript{8} When the model is a
logistic regression, the most basic form of such updating involves
modifying the intercept of the model to account for difference in
prevalence between the source and target
populations.\textsuperscript{1},\textsuperscript{9},\textsuperscript{10}
Because in a logistic regression model, changing the intercept is
equivalent to applying an odds ratio to predicted risks, this method can
be generalized to applying a correcting odds ratio to the outputs of any
risk prediction algorithm, including black-box (e.g., machine learning)
models.\textsuperscript{11}

The transportability of predictive information across populations is an
active area of research in predictive analytics and machine learning. A
recent scoping review categorized methods aimed at developing
transportable markers, or making an existing marker transportable to a
new setting, based on whether they require access to data from the
target population, and whether they are purely data-driven or require
contextual knowledge about associations.\textsuperscript{12} One common
underlying framework in knowledge-driven approaches is causal graphs,
particularly the pioneering work on selection diagrams by Pearl et
al.\textsuperscript{13} Generally, these methods aim to identify and
remove predictors whose association with the outcome varies across
populations.\textsuperscript{14},\textsuperscript{15} Causal diagrams
have recently been used to study how common metrics of model performance
for prognostic and diagnostic markers change across populations with
different case-mix.\textsuperscript{16}

Causal graphs are not the only model for causation. Another is the
sufficient component causes (SCC)
model.\textsuperscript{17},\textsuperscript{18} SCC is a foundational
model in that it establishes fully deterministic relationships between
causes and effects, rather than a representation of statistical
dependencies as in causal graphs. This framework has recently been
applied to explore the biologic plausibility of different link functions
for modeling binary outcomes, resulting in proposals for more
biology-aligned statistical models.\textsuperscript{19}

In this paper, we use a parsimonious SCC model to study the most basic
prediction setup: a binary factor that is used as a marker for the risk
of a binary outcome. As a reference, we formulate this setup for
prognostic markers, where adjustment for outcome prevalence seems to be
a topical issue. This setup is then used to study marker
transportability across populations with varying outcome prevalences.
Our thesis is that by reducing the transportability problem to its basic
constituents in this model, patterns will emerge that can provide
insight into more complex scenarios. This study is entirely theoretical
and simulation-based, and ethics approval was therefore not required.

\subsection{A parsimonious SCC framework for prognostic
markers}\label{a-parsimonious-scc-framework-for-prognostic-markers}

The SCC framework assumes the existence of sets of sufficient causes
that bring about an
event.\textsuperscript{17},\textsuperscript{18},\textsuperscript{20}--\textsuperscript{22}
Within each set, the causes are non-redundant (all elements within the
set are required for the event to happen), but sets can act
independently of each other.

Consider a binary prognostic marker, such as the presence or absence of
Apolipoprotein E (APOE) \(\varepsilon4\) allele, and a binary outcome,
such as the occurrence of Alzheimer's disease.\textsuperscript{23} While
the APOE \(\varepsilon4\) allele is associated with increased
Alzheimer's risk, it is not a definitive marker: neither its absence
rules out the risk, nor does its presence guarantee that cognitive
decline will occur. The fact that neither positive nor negative marker
values are definitive indicates that there are at least two other
mechanisms at play. On one hand, an APOE-\(\varepsilon4\)-positive
individual must experience some other, key events that ultimately lead
to cognitive decline -- explaining why cognitive decline can also occur
via pathways independent of the APOE \(\varepsilon4\) allele --
explaining why many patients with Alzheimer's disease are APOE
\(\varepsilon4\)-negative.

We now formalize a minimal causal setup for the relationship between a
marker value and an outcome. We consider a binary prognostic marker
\(T\) for a binary outcome \(D\). For this marker to be informative but
not definitive, at least two latent variables (or switches) must be
present that can cause false-negative and false-positive responses. We
model these switches as follows:

\begin{itemize}
\item
  The latent binary variable \(U\) represents universally required
  causes -- for example factors that cause subclinical cognitive decline
  to progress to Alzheimer's disease. The absence of \(U\) is
  responsible for false-positive marker values.
\item
  The latent binary variable \(V\) represents all alternative causes --
  for example pathways related to the effect of tobacco smoking, which
  increases the risk of Alzheimer's disease even among those without an
  APOE \(\varepsilon4\) allele.\textsuperscript{24} The presence of
  \(V\) is responsible for false-negative marker values.
\end{itemize}

We assign the values of 1 and 0, respectively, to the `on' and `off'
status of each of these switches. For brevity of notation, a plain
character refers to the `on' value (e.g., \(U\) for \(U=1\)), and its
dot-accented form to the negated value (e.g., \(\dot{U}\) for \(U=0\)).
The above setup results in the following outcome-generating process for
prognostic markers:

\[
D=(T \land U) \lor (V \land U)=(T \lor V) \land U,
\]

where \(\land\) and \(\lor\) are logical AND and OR, respectively.

Given that \(T\), \(U\), and \(V\) are all binary, a population is made
up of 8 subgroups. As \(U\) and \(V\) are latent variables, the observed
properties of the marker in a population are manifested in terms of
\(P(T,D)\), i.e., the two-by-two (contingency) table of marker by
outcome status probabilities. We represent the contingency table by the
sequence \({TP, FP, FN, TN}\), where \(TP=P(T=1,D=1)\),
\(FP=P(T=1,D=0)\), \(FN=P(T=0,D=1)\), \(TN=P(T=0,D=0)\) are,
respectively, true positive, false positive, false negative, and true
negative probabilities.

Our interest is in the transportability of four fundamental performance
characteristics of binary markers: positive predictive value
(\({PPV}=P(D=1|T=1)=TP/(TP+FP)\)), negative predictive value
(\({NPV}=P(D=0|T=0)=TN/(FN+TN)\)), sensitivity
(\({SE}=P(T=1|D=1)=TP/(TP+FN)\)), and specificity
(\({SP}=P(T=0|D=0)=TN/(FP+TN)\)). In particular, we consider various
transportability methods that use the knowledge of outcome prevalence in
the target population. By convention, PPV and NPV are referred to as
predictive values, and SE and SP as accuracy metrics.

\begin{figure}[htbp]
\centering
\begin{center}
\begin{tabular}{p{0.45\linewidth} p{0.45\linewidth}}
\textbf{(A) Reference setup} & \textbf{(B) Symmetrical setup} \\[0.5em]
\hline
\\[1em]
\(D = (T \lor V) \land U\) & \(D = (T \land V) \lor U\)
\\[1em]

\includegraphics[width=0.75\linewidth,keepaspectratio]{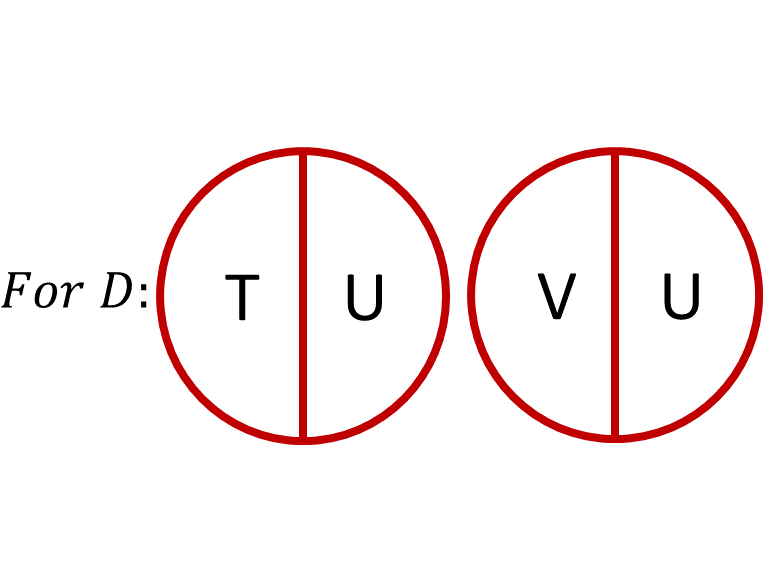} &
\includegraphics[width=0.75\linewidth,keepaspectratio]{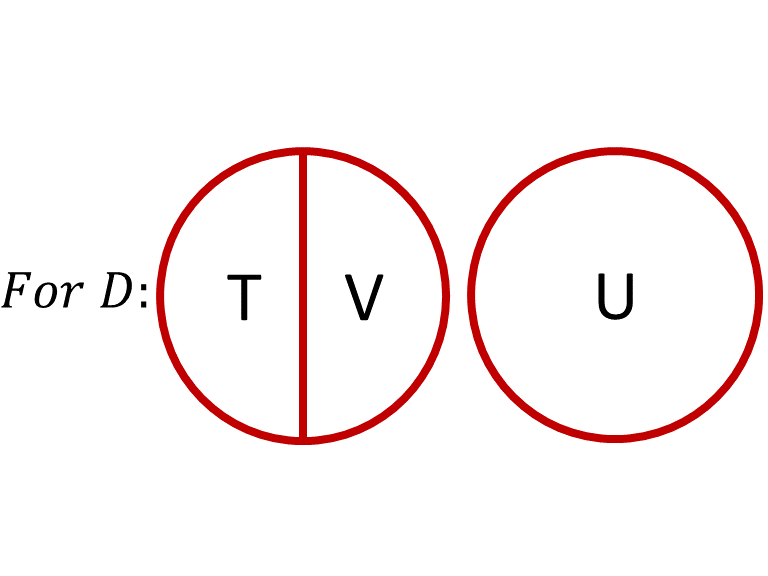} \\[1.5em]

\includegraphics[width=0.4\linewidth,keepaspectratio]{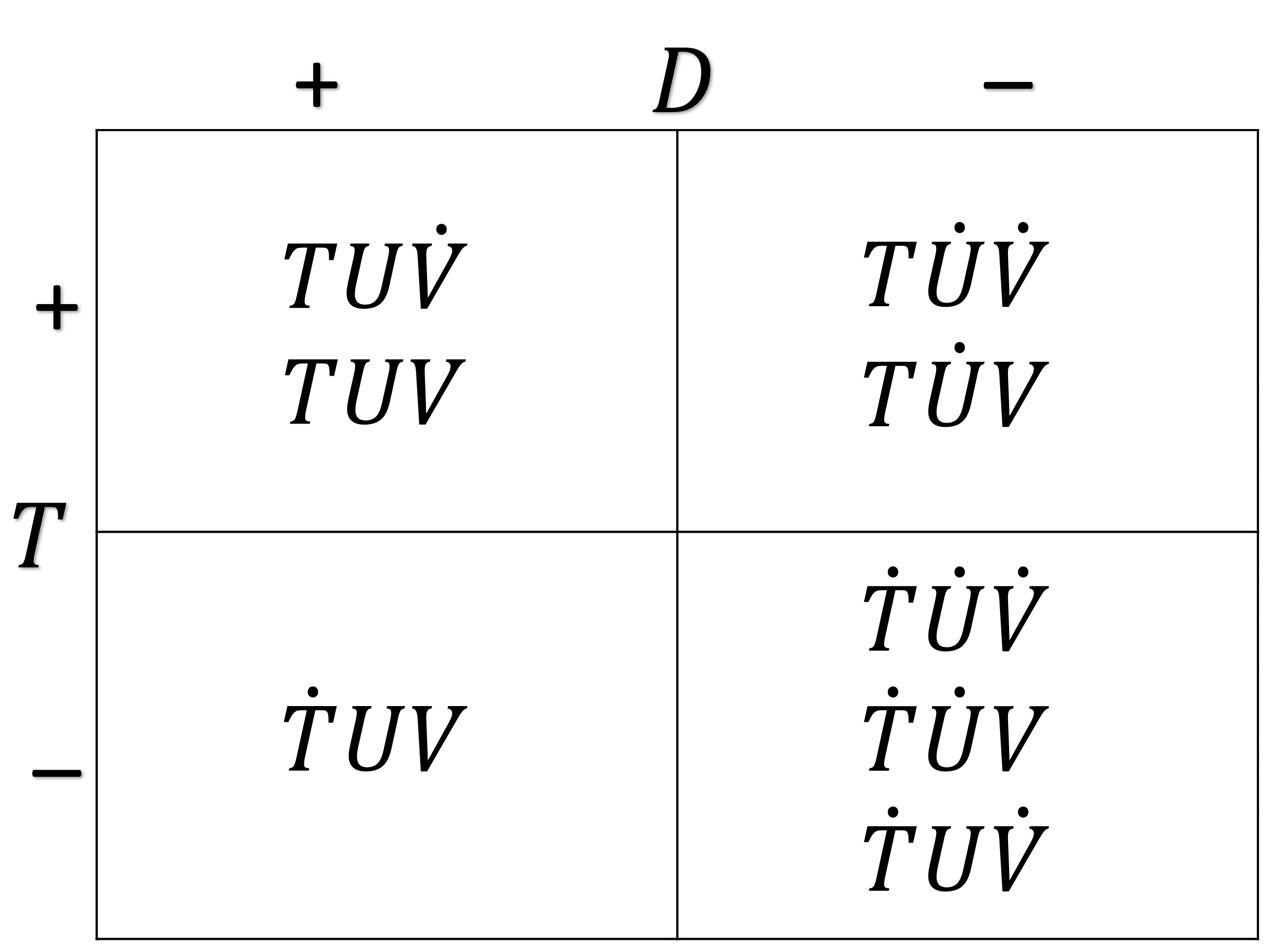} &
\includegraphics[width=0.4\linewidth,keepaspectratio]{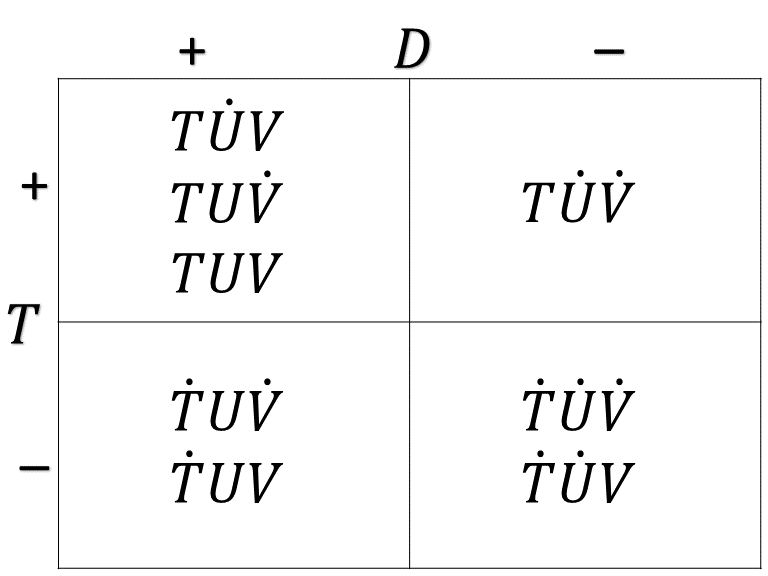} \\[1.5em]
\hline
\end{tabular}
\end{center}
\caption{Minimal configurations for a prognostic marker: the reference (left) and its symmetrical (right) setup, and the resulting contingency tables. Top: logic equation; middle: SCC diagram; bottom: the resulting contingency table. Subgroups are defined as the combination of causes such that the product evaluates to 1; for example, $T\dot{U}V$ is the subgroup where $T=1$, $U=0$, and $V=1$.}
\label{fig:symmetries}
\end{figure}

Before proceeding, we note that one can create an equally parsimonious
SCC setup by swapping the logical AND and OR in the above equation,
resulting in the setup \(D=(T \land V) \lor U\). However, this is
mathematically symmetrical to our reference setup, as the complementary
marker (whose positive and negative results are swapped) in this setup
is a marker for not experiencing the outcome:
\(\dot{D}=(\dot{T} \lor \dot{V}) \land \dot{U}\). Because of this, any
pattern we observe for PPV in the main formulation is also observed for
NPV in this alternative formulation, albeit in the opposite direction,
and vice versa. A similar symmetry exists between SE and SP in these
setups. Figure \ref{fig:symmetries} shows these minimum SCC setups and
the resulting contingency tables (\(P(T,D)\)).

\subsection{Mapping between a contingency table and cause
probabilities}\label{mapping-between-a-contingency-table-and-cause-probabilities}

If the only information available regarding marker performance is a
contingency table, the full joint distribution of the underlying causes
is unidentifiable. Specifying such a joint distribution requires seven
degrees of freedom, whereas the contingency table provides only three.
However, for the purpose of studying performance metrics derived from
these tables, it is sufficient to assume stochastic independence among
the three cause groups. Under this assumption, each cause group follows
a Bernoulli distribution governed by a single parameter, which is its
prevalence. Since three independent Bernoulli distributions require
exactly three degrees of freedom, this setup is identifiable and
possesses a unique (1:1) mapping to the observed table, as is
constructively shown below. This 1:1 mapping demonstrates that any
dependence structure producing a specific contingency table can be
represented by a simplified structure based on stochastic independence
that yields the same results.

We use the notations \(P_T, P_U\), and \(P_V\) to specify the prevalence
of the three cause groups in a population. The 1:1 mapping can be
formulated as follows:

\begin{itemize}
\item
  From \{\(TP, FP, FN, TN\)\} to \{\(P_T, P_U, P_V\)\}:
  \[P_T=TP+FP, \quad P_U=TP/P_T, \quad P_V=FN/[(1-P_T)P_U]\]
\item
  From \{\(P_T,P_U,P_V\)\} to \{\(TP, FP, FN, TN\)\}:
  \[TP=P_T P_U, \quad FP=P_T(1-P_U), \quad FN=(1-P_T)P_U P_V, \quad TN=1-TP-FP-FN.\]
\end{itemize}

As a numerical example, consider a population where \(P_T=\) 0.250,
\(P_U=\) 0.300, and \(P_V=\) 0.150. This results in a prevalence of
0.109, SE of 0.690, SP of 0.804, PPV of 0.300, and NPV of 0.955 (top row
of Table \ref{tbl:example}). These values are chosen to roughly
correspond to the carrier prevalence of APOE \(\varepsilon4\) and
lifetime risk of Alzheimer's disease;\textsuperscript{23} the
prevalences of the universal and alternative causes are illustrative
only. We are interested in transporting this marker to a new population
whose true cause probabilities are \(P_T=\) 0.300, \(P_U=\) 0.400, and
\(P_V=\) 0.200. Outcome prevalence in this population is 0.176, which is
61.8\% higher than in the source population. The true marker performance
metrics in the target population are provided in the second row of Table
\ref{tbl:example}. Assume we do not know the true cause probabilities in
this population but know the outcome prevalence. We will use this simple
numerical example as a case study on various ways a marker can be
transported.

\begin{table}[htbp]
\centering
\begin{threeparttable}
\caption{Numerical example of different marker transportation approaches for the running example}
\label{tbl:example}
\begin{tabular}{ccccccc}
\hline
\textbf{Description} & \{ $P_T, P_U, P_V$ \} & Contingency table\textsuperscript{a} & SE & SP & PPV & NPV \\
\hline
Source population & \{0.250,0.300,0.150\} & \{0.075,0.175,0.034,0.716\} & 0.690 & 0.804 & 0.300 & 0.955 \\ 
Target population & \{0.300,0.400,0.200\} & \{0.120,0.180,0.056,0.644\} & 0.682 & 0.782 & 0.400 & 0.920 \\ 
By predictive values & \{0.300,0.300,0.150\} & \{0.090,0.210,0.031,0.668\} & 0.741 & 0.761 & 0.300 & 0.955 \\ 
By accuracy & \{0.283,0.429,0.178\} & \{0.121,0.162,0.055,0.662\} & 0.690 & 0.804 & 0.429 & 0.924 \\ 
Proportional-odds & \{0.324,0.382,0.203\} & \{0.124,0.201,0.052,0.623\} & 0.703 & 0.757 & 0.382 & 0.923 \\ 
\hline
\end{tabular}
\footnotesize SE: sensitivity; SP: specificity; PPV: positive predictive value; NPV: negative predictive value
\begin{tablenotes}[flushleft]
\footnotesize
\item[a] Contingency table presents the sequence TP (true positive), FP (false positive), FN (false negative), TN (true negative). The top two rows pertain to the true source and target populations; the remaining rows describe the contingency table and performance metrics as implied by each transportation method.
\end{tablenotes}
\end{threeparttable}
\end{table}

\subsection{Two common methods of
transportability}\label{two-common-methods-of-transportability}

Ultimately, using a marker for decision-making entails interpreting its
predictive values, i.e., \(P(D|T)\). Different transportation methods
construct these predictive values by combining different pieces of
information from the source and target populations.

\subsubsection{Transportation by predictive
values}\label{transportation-by-predictive-values}

Transportation by predictive values refers to scenarios where predictive
values (\(P(D|T)\)) in the target population (t) are considered to be
the same as in the source population (s):

\[ PPV^{(t)}=PPV^{(s)} \text{ and } NPV^{(t)}=NPV^{(s)},\] where the
superscripts \(^{(t)}\) and \(^{(s)}\) indicate the values of the
corresponding metric in the target and source populations, respectively.

Given that \(P(T,D)=P(T)P(D|T)\), this transportability mode means the
contingency table in the target population is determined by \(P(T)\) in
that population, combined with predictive values from the source
population. This approach is more common for prognostic markers, a
typical example being risk scoring tools as functions that directly
return an estimate of \(P(D=1|T)\) for someone with marker value \(T\).
For a binary marker, this risk equation can be written as
\(P(D=1|T)=(1-{NPV})+[{PPV}-(1-{NPV})]T\). In our numerical example,
presenting prognostic information as a risk equation would result in
\(P(D=1 | T)=0.045+0.255T\).

\subsubsection{Transportation by
accuracy}\label{transportation-by-accuracy}

In transportation by accuracy, we assume that \(P(T|D)\), i.e., the SE
and SP in the target population, remain the same as in the source
population: \[ SE^{(t)}=SE^{(s)} \text{ and } SP^{(t)}=SP^{(s)}.\]

Given that \(P(T,D)=P(D)P(T|D)\), this transportability mode means the
contingency table in the target population is determined by the outcome
prevalence (\(P(D)\)) in that population, combined with SE and SP from
the source population. This approach is the \emph{modus operandi} for
binary diagnostic markers, where the Bayes' rule is used to combine
pre-test probability with marker accuracy estimates to derive the
post-test probability of the
outcome.\textsuperscript{25},\textsuperscript{26} Without any
information that would distinguish the individual under evaluation, the
pre-test probability is taken to be disease prevalence in the target
population, which is equal to this mode of transportation.

Returning to our running example, transportation by predictive values
preserves PPV and NPV in the target population, but will result in SE of
0.741 and SP of 0.761. If we consider transportation by accuracy, which
uses our knowledge of outcome prevalence in the target population, we
will arrive at the PPV and NPV values of, respectively, 0.429 and 0.924.

The contingency table and marker performance metrics implied by the
above-mentioned transportability methods are provided in Table
\ref{tbl:example} (third and fourth rows).

\paragraph{Prevalence-adjustment for risk
equations:}\label{prevalence-adjustment-for-risk-equations}

A common way that prognostic markers are transported is in the format of
a risk equation for \(P(D|T)\), either explicitly, as is the case in
regression-based prediction models, or implicitly, as in black-box
(e.g., machine learning) models. A familiar modeling framework for
binary outcomes is the logistic regression. There, the logit function
(\(\mbox{logit}(x)=\log(\frac{x}{1-x})\)) is used as link function
connecting marker value to outcome probability. Applying this function
to both sides of the risk equation
\(P(D=1|T)=(1-{NPV})+[{PPV}-(1-{NPV})]T\), we have

\[
\mbox{logit}(P(D=1|T))=\mbox{logit}(P(D=1))+\log\left(\frac{P(T|D=1)}{P(T|D=0)}\right).
\]

The last term on the right-hand side is the likelihood ratio (LR) of the
marker between the diseased and non-diseased groups. The first term on
the right-hand side is the logit of prevalence, which is not a function
of marker value. These derivations indicate that the practice of
prevalence-adjustment by the odds ratio of prevalence between the source
and target populations is equivalent to the application of Bayes'
theorem. This approach changes the locus of transportability from
predictive values to the LR. For binary markers, the LR is defined at
two values. For \(T=0\) it is \((1-SE)/SP\) (aka negative LR), and for
\(T=1\) it is \(SE/(1-SP)\) (aka positive LR). For the marker to be
transportable, both these LRs need to remain constant across
populations, which will be the case if and only if both SE and SP remain
constant. Thus, such prevalence-adjustment is equivalent to
transportation by accuracy metrics.

\subsection{Under what conditions are performance metrics
transportable?}\label{under-what-conditions-are-performance-metrics-transportable}

The above reasoning shows why adjustment for prevalence is generally
expected to improve transportability, as the conventional wisdom is that
SE and SP, being defined within the diseased and non-diseased groups,
are less dependent on prevalence.\textsuperscript{27} While this might
be intuitive as a general observation, under the SCC framework, none of
these metrics are truly intrinsic. Rather, they emerge as properties of
the distribution of cause groups, and their transportability depends on
how this distribution varies across populations. One might be able to
create population-generating mechanisms where a given subset of these
metrics remains constant while others vary. However, our parsimonious
framework helps us examine the plausibility of such mechanisms. Table
\ref{tbl:conditions} shows conditions for the distribution of cause
groups across populations under which each of the four metrics remains
constant (and therefore transportable) for a prognostic marker. These
conditions are derived by mapping transportability requirements
(e.g.~equivalence of SE and SP between the two populations) to cause
group probabilities (an exemplary derivation for PPV is provided in the
footnote of the table).

\begin{table}[htbp]
\centering
\begin{threeparttable}
\caption{Conditions for transportability of metrics of marker performance from a source (s) to a target (t) population}
\label{tbl:conditions}
\begin{tabular}{c >{\centering\arraybackslash}p{0.7\textwidth}}
\hline
\textbf{Metric} & \textbf{Transportability condition} \\
\hline
${PPV}$\textsuperscript{a} & $P_U^{(t)} = P_U^{(s)}$ \\
${NPV}$ & $P_U^{(t)} P_V^{(t)} = P_U^{(s)} P_V^{(s)}$ \\
${SE}$  & $P_V^{(t)}(1-P_T^{(t)})/P_T^{(t)} = P_V^{(s)}(1-P_T^{(s)})/P_T^{(s)}$ \\
${SP}$  & $P_T^{(t)}(1-P_U^{(t)})/[(1-P_T^{(t)})(1-P_U^{(t)} P_V^{(t)})] = P_T^{(s)}(1-P_U^{(s)})/[(1-P_T^{(s)})(1-P_U^{(s)} P_V^{(s)})]$ \\
\hline
\end{tabular}
\footnotesize SE: sensitivity; SP: specificity; PPV: positive predictive value; NPV: negative predictive value
\begin{tablenotes}[flushleft]
\footnotesize
\item[a] As an example of derivations, consider PPV. Its transportability means $P(D=1|T=1)$ should remain the same between the two populations. But $P(D=1|T=1)=(P(T=1,U=1,V=0)+P(T=1,U=1,V=1))/P(T=1)=P(T=1,U=1)/P(T=1)=P(U=1|T=1)=P(U=1)$  (the last equality is based on our assumption of the stochastic independence of causes).
\end{tablenotes}
\end{threeparttable}
\end{table}

For the PPV to be transportable, the prevalence of universal causes
should remain the same between the source and target populations. For
NPV, transportability requires that the proportion of individuals in
whom both the universal and alternative causes are present should remain
the same between the two populations. These conditions are both
satisfied if \(P_U\) and \(P_V\) are stable across populations. In this
scenario, the entirety of variation in outcome prevalence is
attributable to variation in \(P_T\). In our Alzheimer's example, this
would mean populations vary in Alzheimer's disease prevalence only
because they differ in the APOE \(\varepsilon4\) carrier rate. For many
outcomes, this assumption is unrealistic (e.g., many factors contribute
to the variability in Alzheimer's prevalence\textsuperscript{24}). In
comparison, conditions for the transportation by accuracy are more
complicated. For SE, transportability requires the ratio of the
prevalence of alternative causes (\(P_V\)) over the odds of marker
positivity (\(P_T\)) to remain constant. For SP, it does not seem
possible to specify a simple model on how cause groups should vary to
ensure transportability. Overall, transportation by accuracy metrics for
prognostic markers amounts to placing a very specific set of conditions
on how the three groups of causes vary across populations.

\subsection{How do performance metrics vary by prevalence under
different population-generating
mechanisms?}\label{how-do-performance-metrics-vary-by-prevalence-under-different-population-generating-mechanisms}

In our simple causal framework, the degree of transportability of a
marker by a given performance metric depends on how the prevalence of
cause groups vary across populations. Different population-generating
mechanisms can result in different relationships between performance
metrics and outcome prevalence. In this section we visualize how this
relationship changes under various population-generating mechanisms.
Taking the source population of our numerical example as the baseline,
we modeled the following scenarios: when populations vary only in the
prevalence of one of the cause groups; when they vary in the prevalence
of two of the three cause groups; and when they vary in the prevalence
of all three cause groups. For simplicity, when multiple cause groups
vary, we assume they change by the same degree on the odds ratio scale.

\begin{figure}[!htbp]
\centering

\newcommand{\panel}[2]{%
  \begin{minipage}[t]{0.38\textwidth}%
    \raggedright\textbf{(#1)}\par\vspace{0.1em}%
    \includegraphics[width=\linewidth]{#2}%
  \end{minipage}}

\begin{center}
\begin{tabular}{cc}
\panel{A}{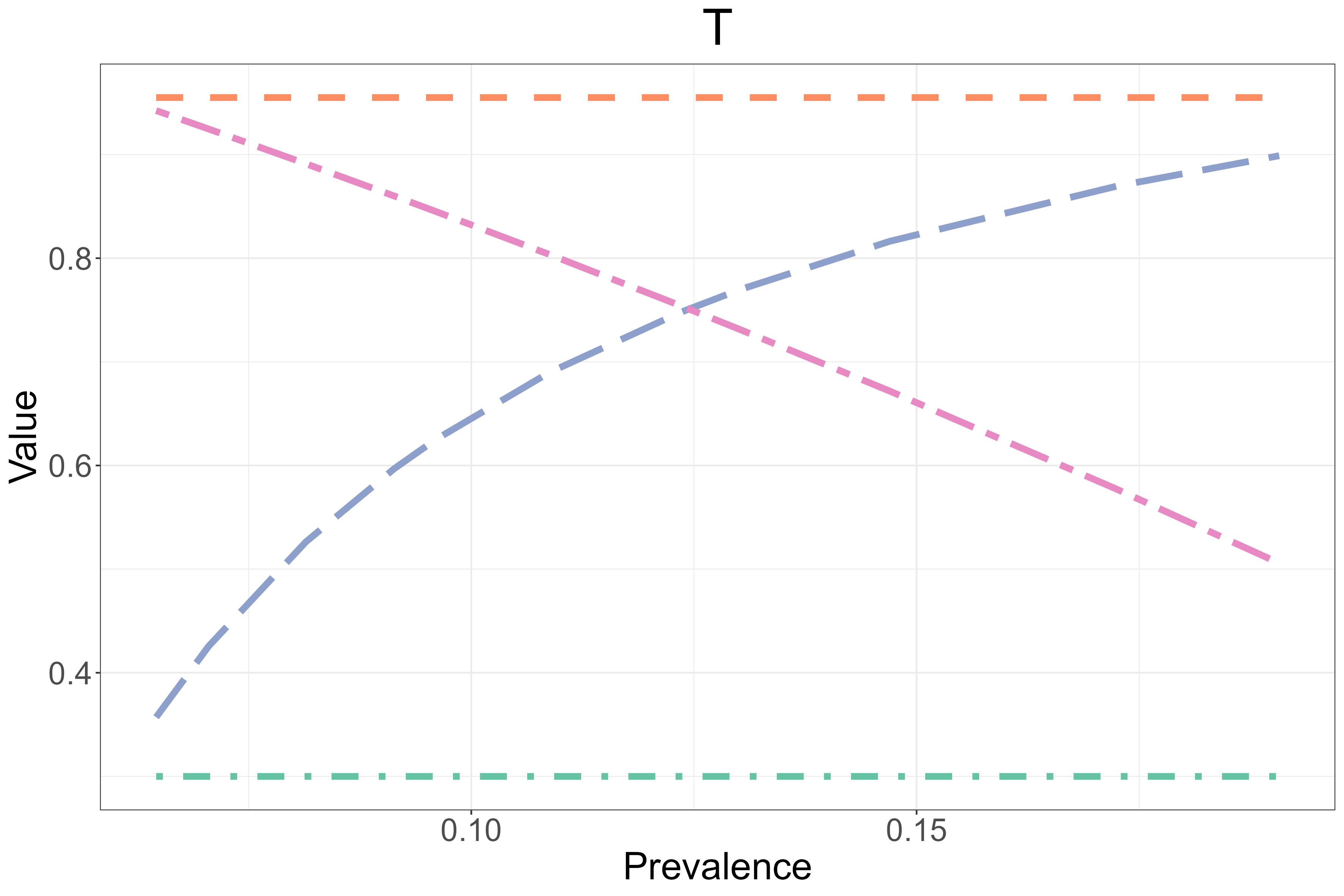}   & \panel{B}{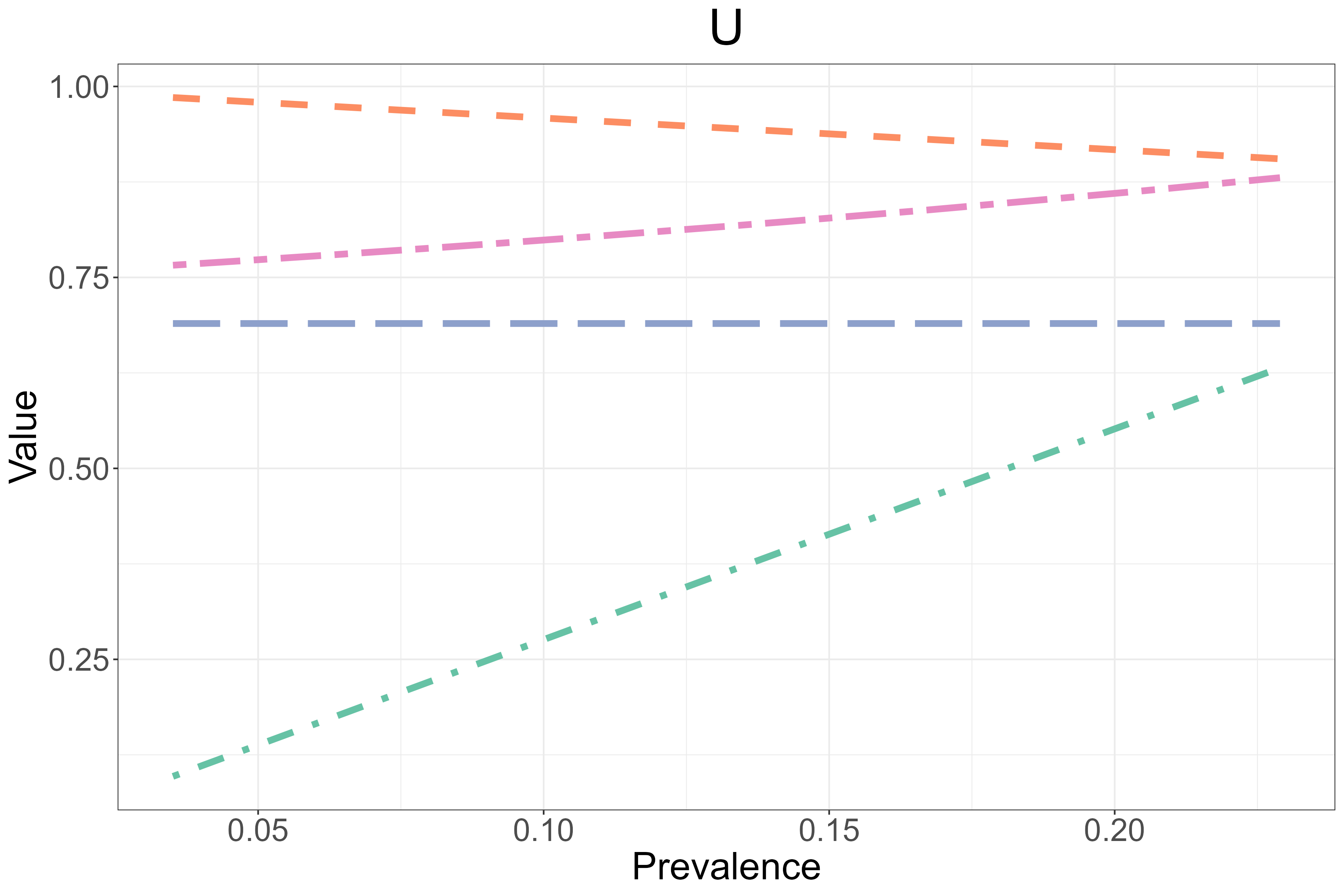}  \\[0.3em]
\panel{C}{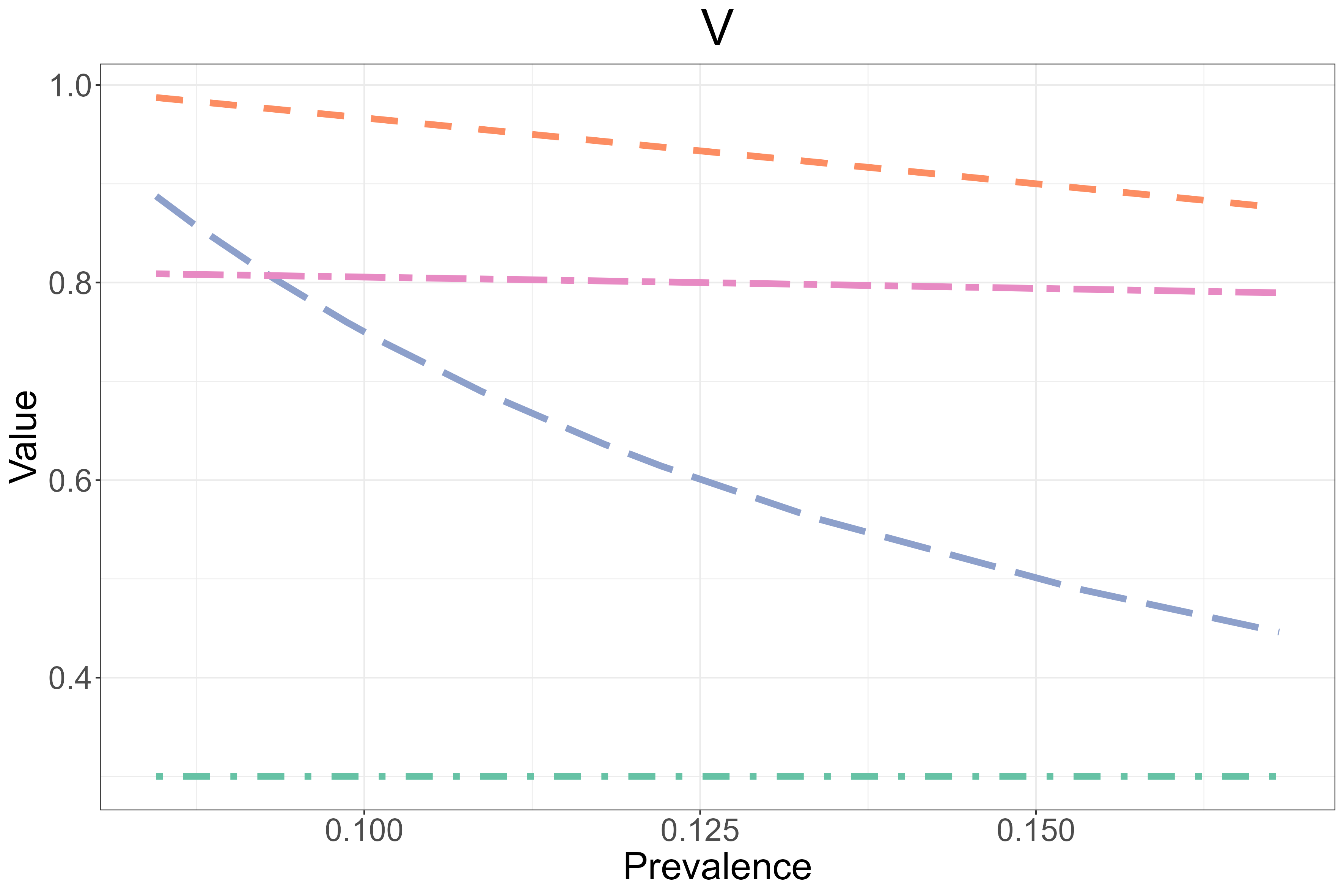}   & \panel{D}{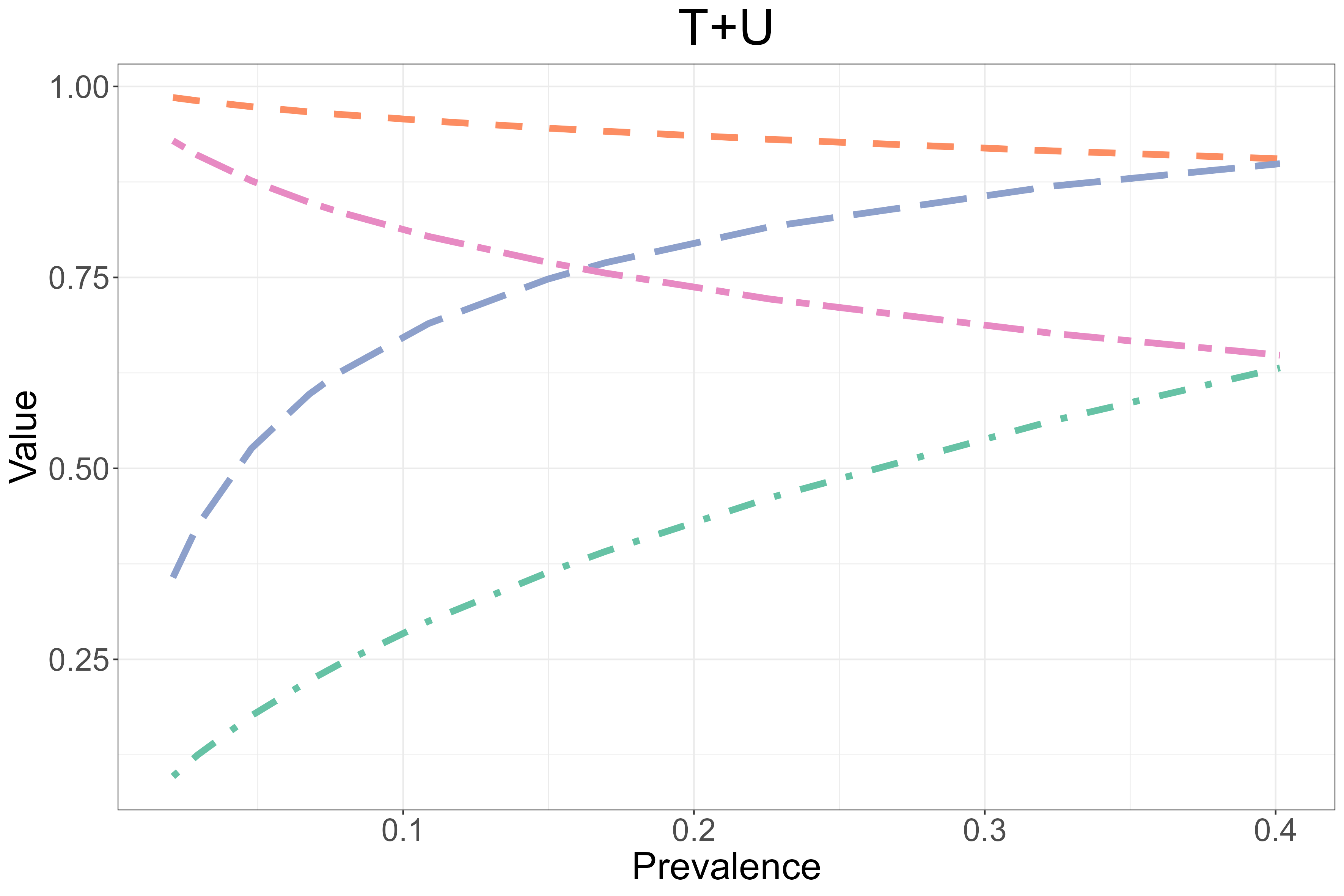} \\[0.3em]
\panel{E}{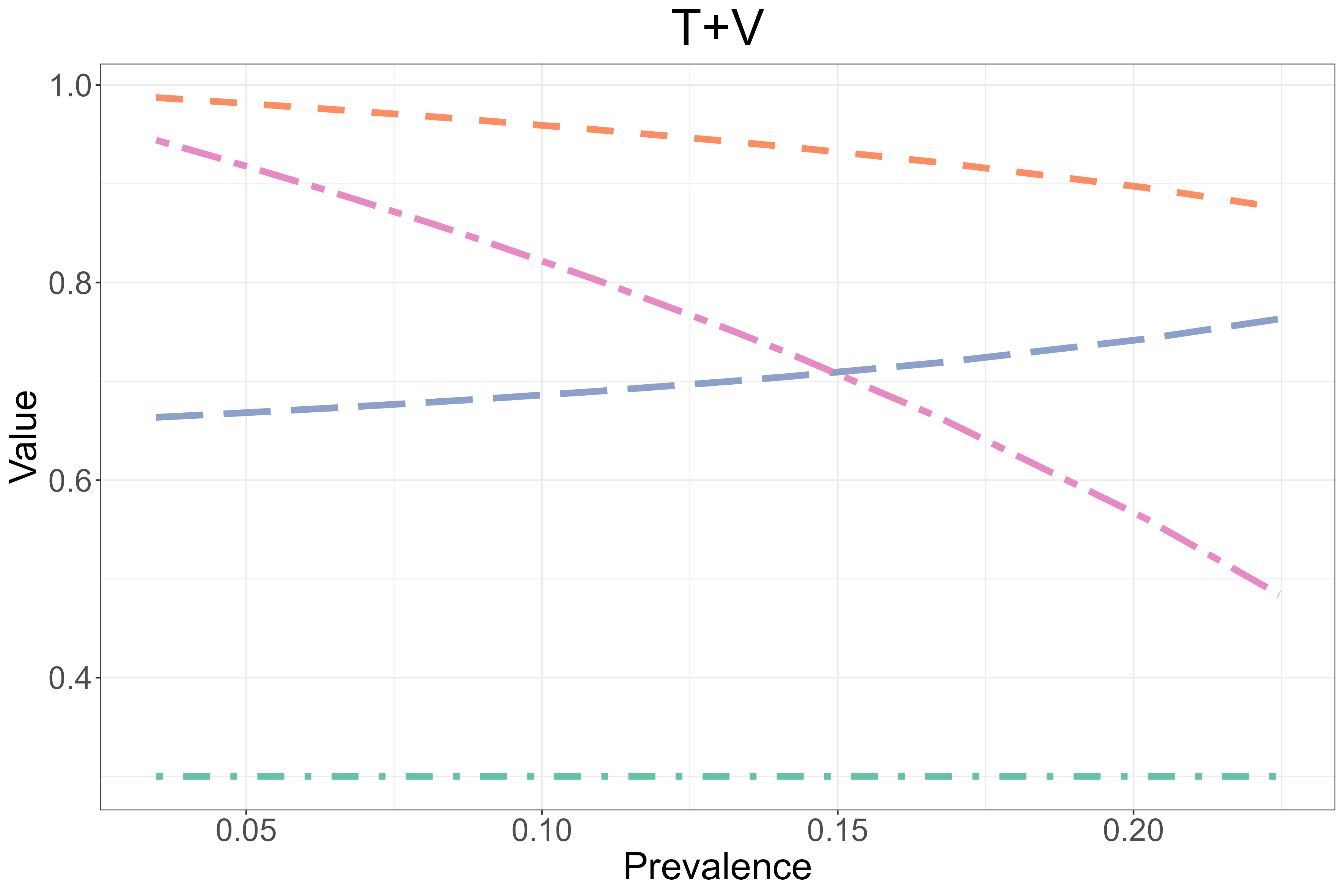}  & \panel{F}{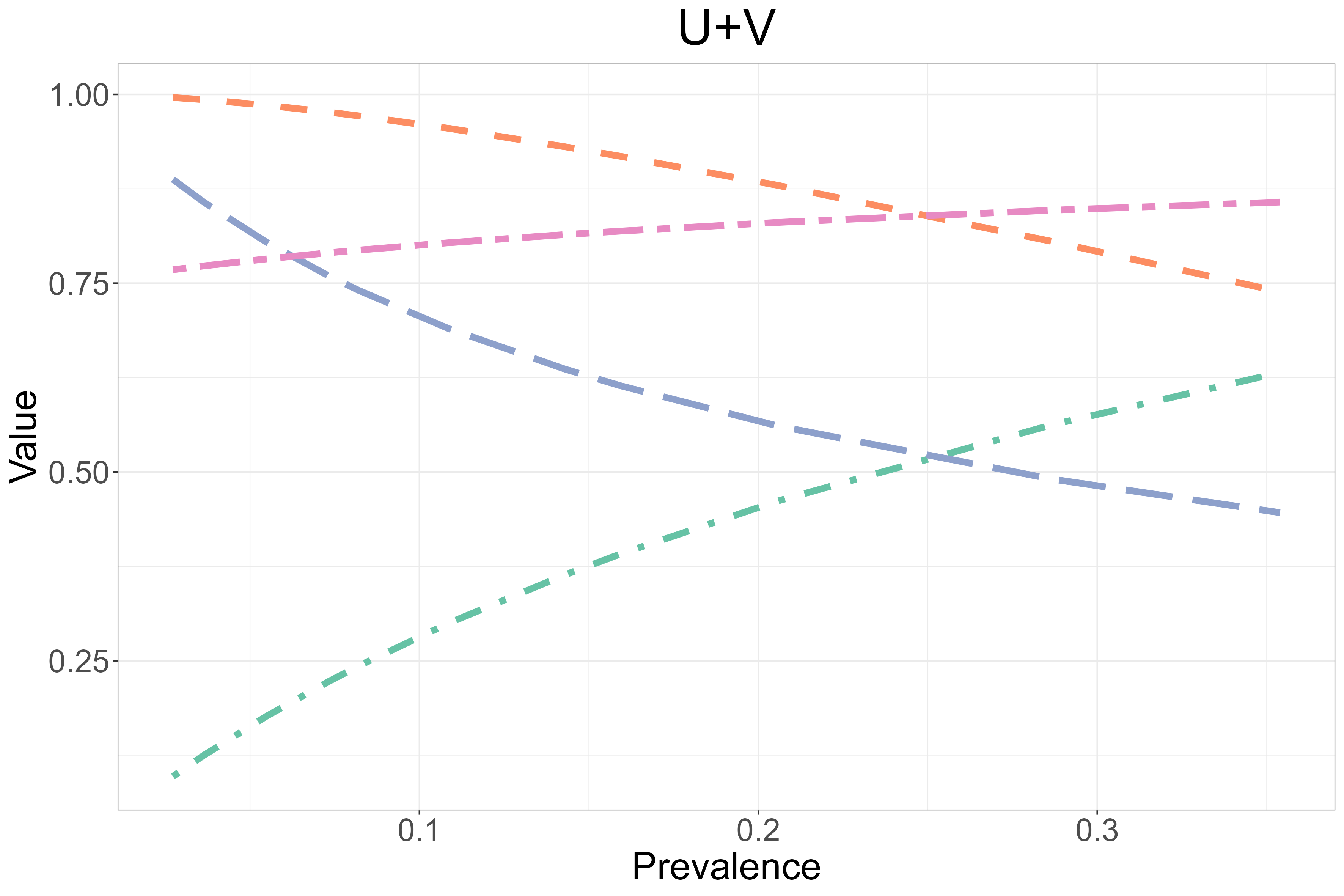} \\[0.3em]
\panel{G}{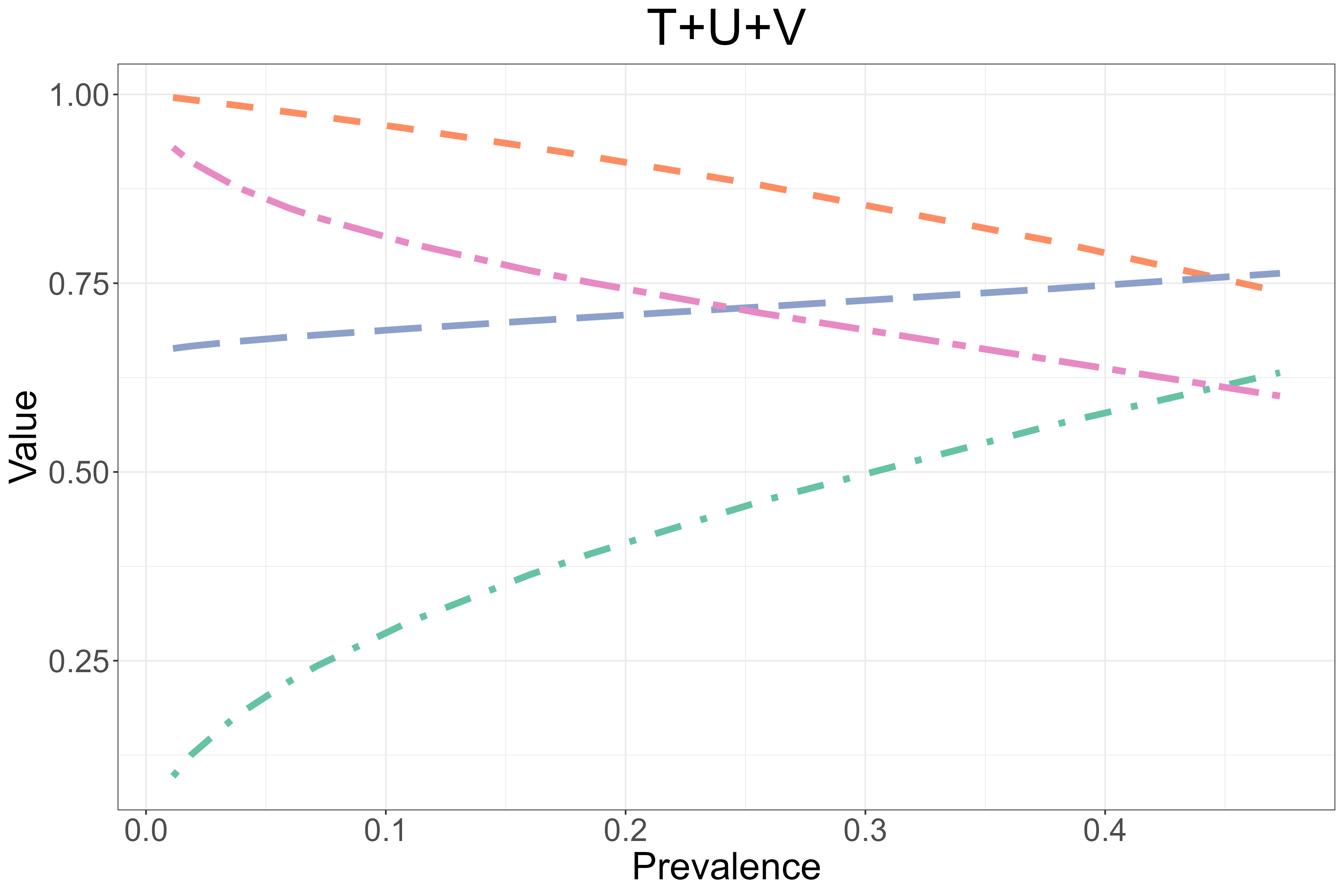} &
\begin{minipage}[t]{0.38\textwidth}%
  \vspace{0pt}%
  \includegraphics[width=\linewidth]{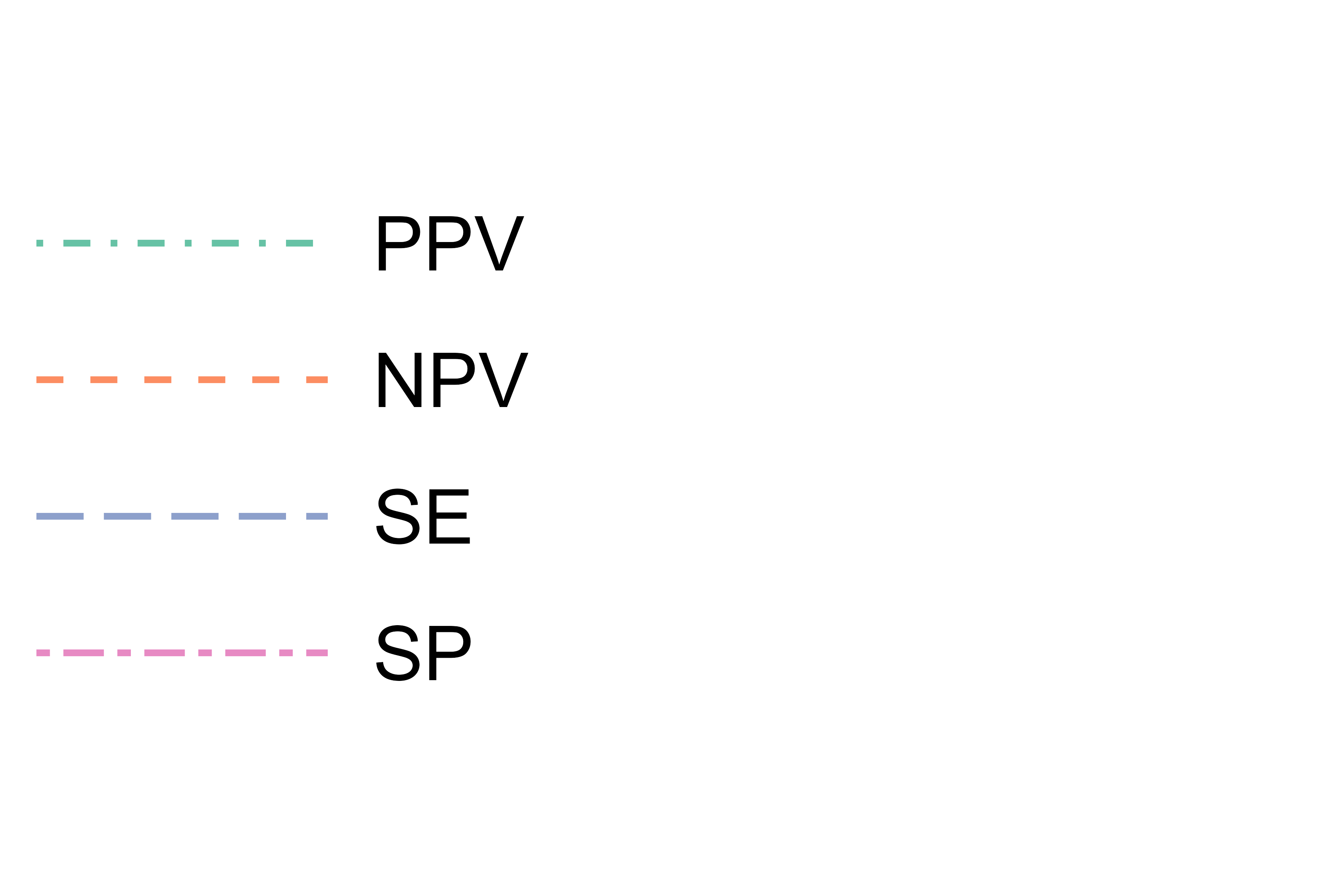}%
\end{minipage} \\
\end{tabular}
\end{center}
\caption{Relationship between prevalence and marker performance metrics under various population-generating mechanisms for a prognostic marker. Panels show variation in: (A) $P_T$ only; (B) $P_U$ only; (C) $P_V$ only; (D) $P_T$ and $P_U$; (E) $P_T$ and $P_V$; (F) $P_U$ and $P_V$; and (G) all three cause groups. SE: sensitivity; SP: specificity; PPV: positive predictive value; NPV: negative predictive value.}
\label{fig:by_prevalence}
\end{figure}

Results are provided in Figure \ref{fig:by_prevalence}. Under \(T\)-only
variation, PPV and NPV remained transportable. This is compatible with
the transportability requirement for predictive values established in
the previous section. Importantly, in all other conditions, PPV and NPV
varied by prevalence. On the other hand, in none of the modeled
scenarios did SE and SP remain unchanged. In fact, SE and SP could vary
in either direction as a function of prevalence, reflecting the more
complex conditions required for their transportability.

\subsection{Proportional-odds transportability
method}\label{proportional-odds-transportability-method}

This framework enables us to express our beliefs on how causal pathways
vary across populations, resulting in algorithms that are transparent
and explicit in their underlying assumptions. Methods based on causal
graphs provide specific solutions for marker transportability if causal
associations between specific factors that determine the marker and
outcome values are known. But, what if the only information we have from
a target population is its outcome prevalence? Our parsimonious
framework enables an overall assessment of the plausibility of the
assumptions under various transportability methods even under such a
general case. For transportability by predictive values, one can
question as to why only one set of causes (\(T\), the one we happen to
be measuring) should be responsible for change in prevalence.
Transportability by accuracy is also questionable due to unclear
assumptions it places on the distribution of causes. For this general
case, it might be more reasonable to take a neutral yet transparent
stance about the variations in causes: that the prevalences of all three
cause groups move by the same extent to bring about change in
prevalence. As an implementation of this approach, one can solve for a
common odds ratio that, when applied to \(P_T\), \(P_U\), and \(P_V\) in
the source population, results in a new population that matches the
desired outcome prevalence. Formally, this approach models the
prevalences of cause groups in the target population as follows:

\[ \mbox{logit}(P_i^{(t)}) = \mbox{logit}(P_i^{(s)}) + \log(x), \]

for \(i \in \{T, U, V\}\), and \(x\) being a common odds ratio. This
``proportional-odds'' assumption is a new, distinct transportability
approach that would generate different updated values of marker
performance metrics compared with both conventional methods. Details of
solving for this common odds ratio are provided in Appendix 1. Appendix
2 provides a basic R implementation.

In our numerical example (last row of Table \ref{tbl:example}), we need
to apply an odds ratio of 1.440 to the prevalences of three cause groups
in the source population to match the target population's outcome
prevalence. The contingency table under this method of transportability
will be \{0.124, 0.201, 0.052, 0.623\}. This in turn results in SE of
0.703 and SP of 0.757. The PPV and NPV are, respectively, 0.382 and
0.923.

The principle underlying the proportional-odds assumption is
impartiality: absent other information, we treat all cause groups as
equally responsible for the shift in outcome prevalence from the source
to the target population. In our running example, this can be justified
given that many causes of Alzheimer's disease (including the prevalence
of the APOE \(\varepsilon4\) allele\textsuperscript{28}) do vary across
populations. Where there is reason to believe that one cause group is
more or less responsible, this can be incorporated explicitly. Consider
a rare germline mutation whose carrier prevalence is stable across
populations (an example is germline APC mutation in familial adenomatous
polyposis as a prognostic marker for colorectal
cancer\textsuperscript{29}). Here it is reasonable to assume that
\(P_T\) remains essentially constant across populations, so that
differences in outcome prevalence are driven by differences in
\(\{P_U,P_V\}\). The resulting transportation method differs from
transportation by predictive values, accuracy metrics, and the
proportional-odds method; indeed, all these methods implicitly require
that a change in outcome prevalence be accompanied by a change in marker
positivity. At the other end of the spectrum, consider diagnosed
hypertension as a prognostic marker for cardiovascular disease. Social
determinants of health (poverty, chronic stress, limited access to
preventive care) plausibly raise, in concert, the prevalence of the
marker itself (hypertension), the universal causes for progression to
overt disease (e.g., low-density lipoproteins necessary for vascular
change), and alternative causal pathways (e.g., hyperglycemia, smoking).
When a single upstream gradient drives all three cause groups together,
and in the absence of any extra information on their prevalences, a
common shift on the odds scale is a reasonable first-order assumption,
and the proportional-odds method becomes a defensible default for
transporting the marker across populations with differing cardiovascular
disease burden.

\subsection{Information loss under different transportability
assumptions}\label{information-loss-under-different-transportability-assumptions}

Ultimately, any assumption about how causes vary across populations will
be a simplified version of a complex reality. When transporting a marker
to a new population, the discrepancy between assumed marker performance
under a given transportability assumption and the true marker
performance results in loss of prognostic information. We conducted
brief simulation studies to explore such information loss under simple
population-generating scenarios. In these simulations, we modeled how
the three cause groups actually vary across populations (the ground
truth). We compared the discrepancy between true marker performance in
the target population and its assumed performance under the
above-mentioned transportability methods.

The discrepancy between a true and assumed marker performance can be
measured in different ways, each focusing on certain aspects of
performance (e.g., discrimination, calibration, or prediction error). A
more foundational approach is to measure information loss by the
Kullback-Leibler divergence (\(D_{KL}\)), an information-theoretic
measure of the discrepancy between a true distribution and a candidate
distribution.\textsuperscript{30} \(D_{KL}\) quantifies the additional
number of bits required to encode information from the true distribution
using the candidate distribution, rather than using the true
distribution itself.\textsuperscript{31} In our case, these
distributions are the true contingency table (\(P(T,D)\)) in the target
population versus the one implied by a given transportability method.

We modeled population-generating scenarios similar to those presented in
Figure \ref{fig:by_prevalence}. Because in many realistic settings all
causes are likely to vary, we explored this setup with more depth,
covering scenarios where the prevalences of cause groups had positive
correlation of varying strengths. In each scenario, the variable-cause
probabilities were generated on the logit scale from a multivariate
normal distribution with mean 0, unit variances, and pairwise
correlation coefficients of 0, 0.25, 0.50, 0.75, and 1 representing
varying degrees of correlation among cause probabilities across
populations. We also modeled a `maximum entropy' scenario where the
three cause groups have independent uniform distributions; albeit
unrealistic, this scenario represents maximum theoretical randomness,
and thus places an upper limit on information loss.\textsuperscript{32}

For each scenario representing a given population-generating mechanism,
we simulated 10,000 random pairs of source and target populations.
Within each pair, we applied the following transportation methods:
transportation by predictive values, transportation by accuracy, and
transportation via the proportional-odds assumption. Within each
simulation iteration, each of these strategies results in an assumed
contingency table for the marker in the target population, which we
compared against the simulated true contingency table in terms of KL
divergence (the lower value, the better).

\begin{figure}[ht]
\centering
\begin{center}
\begin{tabular}{cc}
\includegraphics[width=0.65\linewidth, valign=t]{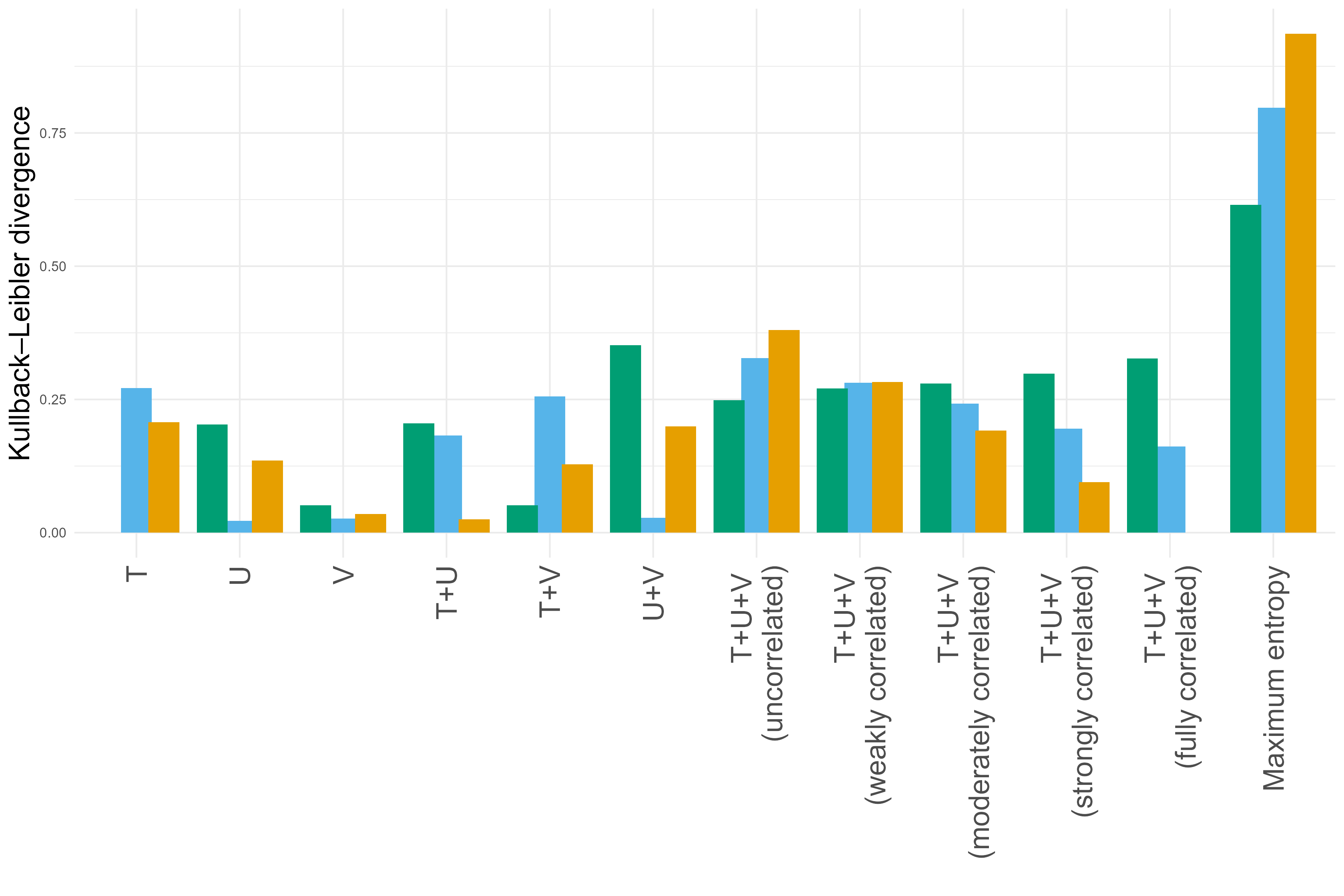} &
\includegraphics[width=0.30\linewidth, valign=t]{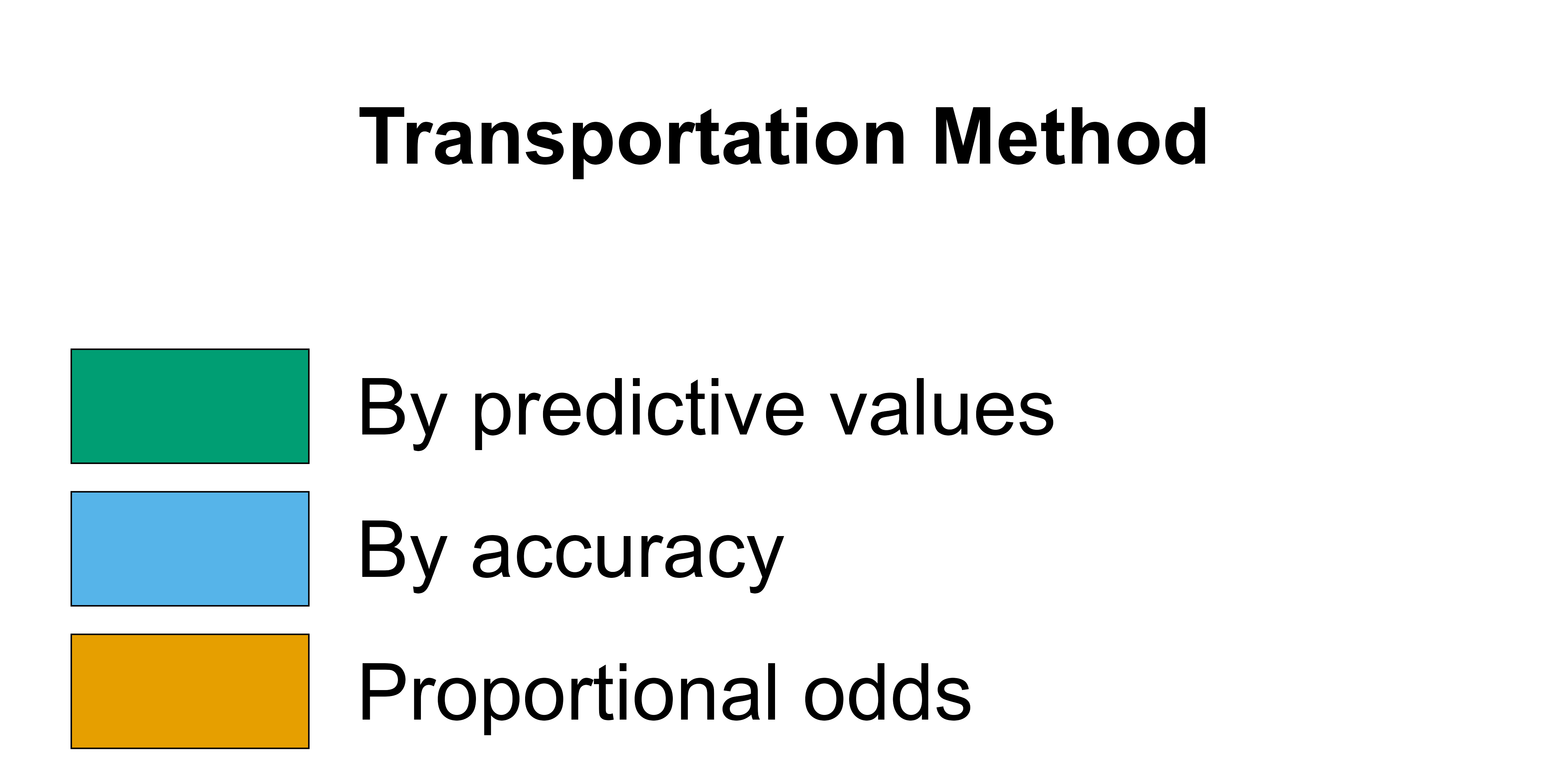} \\
\end{tabular}
\end{center}
\caption{Information loss associated with different methods for transporting a marker under various population-generating scenarios. Kullback-Leibler divergence $(D_{KL})=\sum_{i=1}^4P(a_i)\log_2(P(a_i)/P(b_i))$, with $a_i$ the four cells of the true $P(T,D)$ and $b_i$ the corresponding cells implied by the transportability method. Except in the maximum entropy scenario, the variable component in each scenario had a standard logit-normal distribution. No, weak, moderate, strong, and full correlation correspond to correlation coefficients of 0, 0.25, 0.50, 0.75, and 1 for logit-transformed probabilities.}
\label{fig:simulations}
\end{figure}

Results are provided in Figure \ref{fig:simulations}. Overall, they show
that no transportability method is universally better than others. They
confirm that if the outcome prevalence variation is entirely due to
\(P_T\), PPV and NPV are fully transportable (\(D_{KL}=0\)). However,
under other scenarios, transportation by predictive values resulted in
information loss, sometimes substantially. For transportability by
accuracy, in none of the scenarios was the loss zero, indicating that
none of the modeled scenarios were compatible with stable SE and SP -
again, reflecting the complex requirements for the stability of these
metrics. For the scenarios where the three causes changed together, as
expected, the proportional-odds assumption performed better than both
conventional transportability methods.

\subsection{Discussion}\label{discussion}

We constructed a parsimonious causal framework to study the
transportability of prognostic markers. Several observations from this
theoretical exploration deserve reiterating. First, the common practice
of transporting prognostic markers by predictive values relies on the
strong assumption that variation in outcome prevalence is entirely
attributable to the causes that are captured by the marker. We showed
that the conventional prevalence-adjustment method is the exact
implementation of Bayes' rule, thus changing the locus of
transportability to accuracy (sensitivity and specificity).
Nevertheless, no easily explainable or biologically plausible mechanism
is likely to generate fully transportable accuracy metrics. In our
explored population-generating scenarios, transporting a prognostic
marker by accuracy was not universally better than transporting by
predictive values. However, in scenarios where causes were positively
correlated, which is likely to be common, transporting by accuracy
reduced information loss compared with transportability by predictive
values - this helps justify why prevalence adjustment is generally
expected to improve marker transportability. However, in the same
scenarios, the new `proportional-odds' method of transportation
performed better.

Our numerical results were related to selected ways causes vary across
populations. Still, they should be sufficient to question some common
assumptions and practices, including the practice of advertising
prognostic risk equations without emphasizing their dependence on
case-mix and outcome prevalence, and considering sensitivity and
specificity as intrinsic properties of markers. While these insights
derive from studying binary markers, the core findings may be extendable
to continuous markers and multi-variable risk equations. Transporting
such markers `as is' is equal to assuming that stratum-specific
predictive values (\(P(D|T=x)\)) remain constant. This is equal to
attributing variations in prevalence entirely to variations in \(P_T\).
Consider a multi-variable risk score such as the QRISK3 for
cardiovascular diseases.\textsuperscript{33} Even though this model
includes up to 22 predictors, is it plausible that other causes of
cardiovascular diseases, including environmental exposures, lifestyle
choices, access to and quality of preventive care, are the same across
populations? This assumption is needed for claiming that QRISK3
predictions, developed using primary care UK data, are transportable to
other settings. On the other hand, conventional prevalence-adjustment
would indicate that stratum-specific likelihood ratios would be
transportable (i.e., \(P(T=x|D=1)/P(T=x|D=0)=c\) for all \(x\)). This
assumption will hold if the distributions of predicted risks within
outcome status strata remain the same. This condition places very
specific constraints on how the causes of diseases should vary across
populations.

We focused on prognostic markers given the recent debates on
transportability of prognostic information and the merit of
prevalence-adjustment to improve transportability. The application of
this framework for diagnostic markers is also important and deserves its
own airing. For such markers, the causal direction runs from disease to
test result. As such, an equivalent to our reference setup for a
diagnostic marker would be \(T=(D \lor V) \land U\). From this setup, it
is immediately obvious that transportation by accuracy (the default
approach for diagnostic markers) requires that universal and alternative
causes (\(U\) and \(V\)) remain stable across populations. This strong
requirement can explain why sensitivity and specificity tend to vary
across populations.\textsuperscript{2}--\textsuperscript{4} Further, our
explorations were for when performance metrics are derived from a single
source population. Without knowing how populations differ from each
other, the choice of transportation method will require explicit
assumptions on the distribution of causes (e.g., that they move by the
same value on the odds scale). On the other hand, when contingency
tables from multiple populations are at hand, the relationship between
cause distributions can be learned from the data. This line of reasoning
can result in novel meta-analytic approaches based on modeling
\((P_T, P_U, P_V)\) across populations, offering not just an alternative
to pooling accuracy metrics or predictive
values,\textsuperscript{34},\textsuperscript{35} but a mechanistic
explanation for the heterogeneity that routinely undermines such
meta-analyses. Yet another line of inquiry would be the use of this
method for modeling the joint distribution of multiple marker values - a
common challenge in practice. If it can be assumed that markers have
shared underlying pathways, their joint distribution might provide
additional insight on how those shared pathways change across
populations.

To conclude, we question the contemporary practice of advertising risk
prediction models for outcome risk as transportable. If reliable,
context-specific information is available on the distribution of
predictors and their relationships with the outcome and with each other,
methods based on causal graphs can be used to build transportable models
or to design tailored transportation strategies.\textsuperscript{14} In
contrast, when only general information, such as outcome prevalence or
test positivity rates, is available, the SCC framework can be used to
formulate transportability algorithms that utilize such broad
information. Instead of fixed transportability rules, this framework
offers a foundation for algorithms that reflect our assumptions on how
underlying causes vary, rather than keeping performance metrics, which
are emergent properties of such causes, constant.

\newpage

\subsection{References}\label{references}

\protect\phantomsection\label{refs}

\newpage

\newpage

\section{Appendix (Supplemental
Material)}\label{appendix-supplemental-material}

\subsection{Appendix 1: Transportation by the proportional-odds
assumption}\label{appendix-1-transportation-by-the-proportional-odds-assumption}

Let \{\(TP^{(s)}, FP^{(s)}, FN^{(s)}, TN^{(s)}\)\} be the elements of
the two-by-two contingency table in the source population \(s\). We
would like to transport this marker to a new population \(p\) where we
know the disease prevalence is \(\pi\). Our goal is to construct a
predicted contingency table for this target population, defined by
\{\(TP^{(t)}, FP^{(t)}, FN^{(t)}, TN^{(t)}\)\}.

Step 1. Map from \{\(TP^{(s)}, FP^{(s)}, FN^{(s)}, TN^{(s)}\)\} to
\{\(P_T^{(s)},P_U^{(s)},P_V^{(s)}\)\} (see text).

Step 2. Find the odds ratio \(x\) such that when applied to
\(P_T^{(s)}\), \(P_U^{(s)}\), \(P_V^{(s)}\), results in updated
probabilities \(P_T^{(t)}\), \(P_U^{(t)}\), \(P_V^{(t)}\) that
correspond to the desired prevalence. Note that prevalence in the target
population is
\(\pi=P_U^{(t)}(P_T^{(t)}+P_V^{(t)}-P_T^{(t)} P_V^{(t)})\).

Given that applying an odds ratio \(x\) to probability \(p\) can be
written as the function \(\frac{px}{1-p+px}\), we can write the
prevalence of each cause in the target population as a function of
\(x\):

\(P_T^{(t)}(x)=\frac{P_T^{(s)}x}{1-P_T^{(s)}+P_T^{(s)}x}\) (similar for
\(P_U^{(t)}\) and \(P_V^{(t)}\)).

Thus, we should solve for \(x\) in \(f(x)=\pi\) where \[
f(x)=P_U^{(t)}(x)[P_T^{(t)}(x)+P_V^{(t)}(x)-P_T^{(t)}(x)P_V^{(t)}(x)].
\] We consider non-degenerate scenarios where \(0<P_T^{(s)}<1\),
\(0<P_U^{(s)}<1\), and \(0<P_V^{(s)}<1\). Given that \(f(x)\) is
continuous, \(f(0)=0\), and \(\lim_{x \to \infty}f(x)=1\), there is
always at least one real solution for \(x\). Further, given that

\begin{itemize}
\tightlist
\item
  \(\frac{dP_T^{(t)}(x)}{dx}>0\), \(\frac{dP_U^{(t)}(x)}{dx}>0\), and
  \(\frac{dP_V^{(t)}(x)}{dx}>0\), and
\item
  \(\frac{\partial f(x)}{\partial P_T^{(t)}(x)}>0\),
  \(\frac{\partial f(x)}{\partial P_U^{(t)}(x)}>0\),
  \(\frac{\partial f(x)}{\partial P_V^{(t)}(x)}>0\),
\end{itemize}

\(\frac{df(x)}{dx}>0\) for all \(x\). As such, this solution is unique.
This solution can be found using univariate root finding methods (e.g.,
\emph{uniroot()} in R).

Of note, re-arranging the terms reveals that \(x\) can also be expressed
as the real root of the cubic equation \(ax^3+bx^2+cx+d=0\) with

\begin{align*}
  a &= (1-\pi)P_T^{(s)} P_U^{(s)} P_V^{(s)}, \\
  b &= (3\pi -2) P_T^{(s)} P_U^{(s)} P_V^{(s)} - \pi[P_T^{(s)} P_U^{(s)} + P_T^{(s)} P_V^{(s)} + P_U^{(s)} P_V^{(s)}] + P_T^{(s)} P_U^{(s)} + P_U^{(s)} P_V^{(s)}, \\
  c &= -\pi[3P_T^{(s)} P_U^{(s)} P_V^{(s)} - 2P_T^{(s)}P_U^{(s)} - 2P_T^{(s)}P_V^{(s)} - 2P_U^{(s)}P_V^{(s)} + P_T^{(s)} + P_U^{(s)} + P_V^{(s)}], \\
  d &= -\pi(1-P_T^{(s)})(1-P_U^{(s)})(1-P_V^{(s)});
\end{align*}

which can be solved using standard methods (e.g., \emph{polyroot()} in
R).

\begin{enumerate}
\def\labelenumi{\arabic{enumi}.}
\setcounter{enumi}{2}
\tightlist
\item
  Map the resulting \{\(P_T^{(t)}\), \(P_U^{(t)}\), \(P_V^{(t)}\)\} to
  \{\(TP^{(t)},FP^{(t)},FN^{(t)},TN^{(t)}\)\} (see text).
\end{enumerate}

\newpage

\subsection{Appendix 2: R code for basic implementation of
transportability
methods}\label{appendix-2-r-code-for-basic-implementation-of-transportability-methods}

\footnotesize

This function returns the positive and negative predictive values of a
prognostic marker in the new target population, given the contingency
table in the source population, an estimate of outcome prevalence in the
target population, and a given transportability assumption.

\begin{Shaded}
\begin{Highlighting}[]
\NormalTok{transport\_marker }\OtherTok{\textless{}{-}} \ControlFlowTok{function}\NormalTok{(TP, FP, FN, TN, prev, assumption) \{}
  \ControlFlowTok{switch}\NormalTok{(}
\NormalTok{    assumption,}
    \AttributeTok{pv =}\NormalTok{ \{}
\NormalTok{      PPV }\OtherTok{\textless{}{-}}\NormalTok{ TP }\SpecialCharTok{/}\NormalTok{ (TP }\SpecialCharTok{+}\NormalTok{ FP)}
\NormalTok{      NPV }\OtherTok{\textless{}{-}}\NormalTok{ TN }\SpecialCharTok{/}\NormalTok{ (TN }\SpecialCharTok{+}\NormalTok{ FN)}
\NormalTok{    \},}
    \AttributeTok{ac =}\NormalTok{ \{}
\NormalTok{      SE }\OtherTok{\textless{}{-}}\NormalTok{ TP }\SpecialCharTok{/}\NormalTok{ (TP }\SpecialCharTok{+}\NormalTok{ FN)}
\NormalTok{      SP }\OtherTok{\textless{}{-}}\NormalTok{ TN }\SpecialCharTok{/}\NormalTok{ (TN }\SpecialCharTok{+}\NormalTok{ FP)}
\NormalTok{      PPV }\OtherTok{\textless{}{-}}\NormalTok{ prev }\SpecialCharTok{*}\NormalTok{ SE }\SpecialCharTok{/}\NormalTok{ (prev }\SpecialCharTok{*}\NormalTok{ SE }\SpecialCharTok{+}\NormalTok{ (}\DecValTok{1} \SpecialCharTok{{-}}\NormalTok{ prev) }\SpecialCharTok{*}\NormalTok{ (}\DecValTok{1} \SpecialCharTok{{-}}\NormalTok{ SP))}
\NormalTok{      NPV }\OtherTok{\textless{}{-}}\NormalTok{ (}\DecValTok{1} \SpecialCharTok{{-}}\NormalTok{ prev) }\SpecialCharTok{*}\NormalTok{ SP }\SpecialCharTok{/}\NormalTok{ ((}\DecValTok{1} \SpecialCharTok{{-}}\NormalTok{ prev) }\SpecialCharTok{*}\NormalTok{ SP }\SpecialCharTok{+}\NormalTok{ prev }\SpecialCharTok{*}\NormalTok{ (}\DecValTok{1} \SpecialCharTok{{-}}\NormalTok{ SE))}
\NormalTok{    \},}
    \AttributeTok{po =}\NormalTok{ \{}
\NormalTok{      PT }\OtherTok{\textless{}{-}}\NormalTok{ TP }\SpecialCharTok{+}\NormalTok{ FP}
\NormalTok{      PU }\OtherTok{\textless{}{-}}\NormalTok{ TP }\SpecialCharTok{/}\NormalTok{ PT}
\NormalTok{      PV }\OtherTok{\textless{}{-}}\NormalTok{ FN }\SpecialCharTok{/}\NormalTok{ (}\DecValTok{1} \SpecialCharTok{{-}}\NormalTok{ PT) }\SpecialCharTok{/}\NormalTok{ PU}
\NormalTok{      res }\OtherTok{\textless{}{-}} \FunctionTok{uniroot}\NormalTok{(}
        \ControlFlowTok{function}\NormalTok{(x) \{}
\NormalTok{          PTt }\OtherTok{\textless{}{-}}\NormalTok{ PT }\SpecialCharTok{*}\NormalTok{ x }\SpecialCharTok{/}\NormalTok{ (}\DecValTok{1} \SpecialCharTok{{-}}\NormalTok{ PT }\SpecialCharTok{+}\NormalTok{ PT }\SpecialCharTok{*}\NormalTok{ x)}
\NormalTok{          PUt }\OtherTok{\textless{}{-}}\NormalTok{ PU }\SpecialCharTok{*}\NormalTok{ x }\SpecialCharTok{/}\NormalTok{ (}\DecValTok{1} \SpecialCharTok{{-}}\NormalTok{ PU }\SpecialCharTok{+}\NormalTok{ PU }\SpecialCharTok{*}\NormalTok{ x)}
\NormalTok{          PVt }\OtherTok{\textless{}{-}}\NormalTok{ PV }\SpecialCharTok{*}\NormalTok{ x }\SpecialCharTok{/}\NormalTok{ (}\DecValTok{1} \SpecialCharTok{{-}}\NormalTok{ PV }\SpecialCharTok{+}\NormalTok{ PV }\SpecialCharTok{*}\NormalTok{ x)}
\NormalTok{          PUt }\SpecialCharTok{*}\NormalTok{ (PTt }\SpecialCharTok{+}\NormalTok{ PVt }\SpecialCharTok{{-}}\NormalTok{ PTt }\SpecialCharTok{*}\NormalTok{ PVt) }\SpecialCharTok{{-}}\NormalTok{ prev}
\NormalTok{        \},}
        \AttributeTok{interval =} \FunctionTok{c}\NormalTok{(}\FloatTok{0.01}\NormalTok{, }\DecValTok{100}\NormalTok{)}
\NormalTok{      )}
\NormalTok{      x }\OtherTok{\textless{}{-}}\NormalTok{ res}\SpecialCharTok{$}\NormalTok{root}
      \FunctionTok{message}\NormalTok{(}\StringTok{"Common odds ratio is "}\NormalTok{, x)}
\NormalTok{      PUt }\OtherTok{\textless{}{-}}\NormalTok{ PU }\SpecialCharTok{*}\NormalTok{ x }\SpecialCharTok{/}\NormalTok{ (}\DecValTok{1} \SpecialCharTok{{-}}\NormalTok{ PU }\SpecialCharTok{+}\NormalTok{ PU }\SpecialCharTok{*}\NormalTok{ x)}
\NormalTok{      PVt }\OtherTok{\textless{}{-}}\NormalTok{ PV }\SpecialCharTok{*}\NormalTok{ x }\SpecialCharTok{/}\NormalTok{ (}\DecValTok{1} \SpecialCharTok{{-}}\NormalTok{ PV }\SpecialCharTok{+}\NormalTok{ PV }\SpecialCharTok{*}\NormalTok{ x)}
\NormalTok{      PPV }\OtherTok{\textless{}{-}}\NormalTok{ PUt}
\NormalTok{      NPV }\OtherTok{\textless{}{-}} \DecValTok{1} \SpecialCharTok{{-}}\NormalTok{ PUt }\SpecialCharTok{*}\NormalTok{ PVt}
\NormalTok{    \},}
    \FunctionTok{stop}\NormalTok{(}
      \StringTok{"Assumption should be one of pv (by predictive values), ac (by accuracy), or po (by proportional{-}odds)"}
\NormalTok{    )}
\NormalTok{  )}
  \FunctionTok{c}\NormalTok{(}\AttributeTok{PPV =}\NormalTok{ PPV, }\AttributeTok{NPV =}\NormalTok{ NPV)}
\NormalTok{\}}
\end{Highlighting}
\end{Shaded}

Reproducing the results of the running example. The four cell
probabilities are taken from the source population of Table 1. Outcome
prevalence in the target population is 0.176.

\begin{Shaded}
\begin{Highlighting}[]
\CommentTok{\#Change \textquotesingle{}po\textquotesingle{} (proportional{-}odds) to \textquotesingle{}pv\textquotesingle{} (by predictive values) or \textquotesingle{}ac\textquotesingle{} (by accuracy)}
\FunctionTok{transport\_marker}\NormalTok{(}\AttributeTok{TP=}\FloatTok{0.075}\NormalTok{, }\AttributeTok{FP=}\FloatTok{0.175}\NormalTok{, }\AttributeTok{FN=}\FloatTok{0.03375}\NormalTok{, }\AttributeTok{TN=}\FloatTok{0.71625}\NormalTok{, }\AttributeTok{prev=}\FloatTok{0.176}\NormalTok{, }\AttributeTok{assumption=}\StringTok{"po"}\NormalTok{)}
\end{Highlighting}
\end{Shaded}

\begin{verbatim}
## Common odds ratio is 1.43988375321327
\end{verbatim}

\begin{verbatim}
##       PPV       NPV 
## 0.3816064 0.9226814
\end{verbatim}

\normalsize


\begin{CSLReferences}{0}{1}
\bibitem[\citeproctext]{ref-Morise1996InterceptAdjustment}
\CSLLeftMargin{1. }%
\CSLRightInline{Morise AP, Diamond GA, Detrano R, Bobbio M, Gunel E.
\href{https://doi.org/10.1177/0272989x9601600205}{The effect of
disease-prevalence adjustments on the accuracy of a logistic prediction
model}. Med Decis Making. 1996;16:133--142.}

\bibitem[\citeproctext]{ref-Murad2023SeSpPrev}
\CSLLeftMargin{2. }%
\CSLRightInline{Murad MH, Lin L, Chu H, et al. The association of
sensitivity and specificity with disease prevalence. CMAJ.
2023;195:E925--E931.}

\bibitem[\citeproctext]{ref-Leeflang2009SeSpPrev}
\CSLLeftMargin{3. }%
\CSLRightInline{Leeflang MMG, Bossuyt PMM, Irwig L.
\href{https://doi.org/10.1016/j.jclinepi.2008.04.007}{Diagnostic test
accuracy may vary with prevalence: Implications for evidence-based
diagnosis}. J Clin Epidemiol. 2009;62:5--12.}

\bibitem[\citeproctext]{ref-BRENNER1997SeSpPPVNPVPrevVar}
\CSLLeftMargin{4. }%
\CSLRightInline{Brenner H, Gefeller O.
\href{https://doi.org/10.1002/(sici)1097-0258(19970515)16:9\%3C981::aid-sim510\%3E3.0.co;2-n}{Variation
of sensitivity, specificity, likelihood ratios and predictive values
with disease prevalence}. Stat Med. 1997;16:981--991.}

\bibitem[\citeproctext]{ref-gulati2022}
\CSLLeftMargin{5. }%
\CSLRightInline{Gulati G, Upshaw J, Wessler BS, et al.
\href{https://doi.org/10.1161/CIRCOUTCOMES.121.008487}{Generalizability
of Cardiovascular Disease Clinical Prediction Models: 158 Independent
External Validations of 104 Unique Models}. Circ Cardiovasc Qual
Outcomes. 2022;15:e008487.}

\bibitem[\citeproctext]{ref-Ho2023COPDCPMVariability}
\CSLLeftMargin{6. }%
\CSLRightInline{Ho JK, Safari A, Adibi A, et al.
\href{https://doi.org/10.1016/j.chest.2022.11.041}{Generalizability of
risk stratification algorithms for exacerbations in COPD}. Chest.
2023;163:790--798.}

\bibitem[\citeproctext]{ref-zech2018}
\CSLLeftMargin{7. }%
\CSLRightInline{Zech JR, Badgeley MA, Liu M, Costa AB, Titano JJ,
Oermann EK. \href{https://doi.org/10.1371/journal.pmed.1002683}{Variable
generalization performance of a deep learning model to detect pneumonia
in chest radiographs: A cross-sectional study}. PLoS Med.
2018;15:e1002683.}

\bibitem[\citeproctext]{ref-steyerberg2019}
\CSLLeftMargin{8. }%
\CSLRightInline{Steyerberg EW. Updating for a new setting. In: Clinical
prediction models. 2nd ed. Cham: Springer International Publishing;
2019. p. 399--429.}

\bibitem[\citeproctext]{ref-janssen2008}
\CSLLeftMargin{9. }%
\CSLRightInline{Janssen KJM, Moons KGM, Kalkman CJ, Grobbee DE, Vergouwe
Y. \href{https://doi.org/10.1016/j.jclinepi.2007.04.018}{Updating
methods improved the performance of a clinical prediction model in new
patients}. J Clin Epidemiol. 2008;61:76--86.}

\bibitem[\citeproctext]{ref-sadatsafavi2022}
\CSLLeftMargin{10. }%
\CSLRightInline{Sadatsafavi M, Tavakoli H, Safari A.
\href{https://doi.org/10.1097/ede.0000000000001489}{Marginal versus
conditional odds ratios when updating risk prediction models}.
Epidemiology. 2022;33:555--558.}

\bibitem[\citeproctext]{ref-Meijerink2025MLUpdatingReview}
\CSLLeftMargin{11. }%
\CSLRightInline{Meijerink LM, Dunias ZS, Leeuwenberg AM, et al.
\href{https://doi.org/10.1016/j.jclinepi.2024.111636}{Updating methods
for artificial intelligence--based clinical prediction models: A scoping
review}. J Clin Epidemiol. 2025;178:111636.}

\bibitem[\citeproctext]{ref-ploddi2024}
\CSLLeftMargin{12. }%
\CSLRightInline{Ploddi K, Sperrin M, Martin GP, O'Connell MM.
\href{https://doi.org/10.48550/ARXIV.2412.04275}{Scoping review of
methodology for aiding generalisability and transportability of clinical
prediction models} {[}Internet{]}. 2024.}

\bibitem[\citeproctext]{ref-pearl2011}
\CSLLeftMargin{13. }%
\CSLRightInline{Pearl J, Bareinboim E.
\href{https://doi.org/10.1109/icdmw.2011.169}{Transportability of causal
and statistical relations: A formal approach}. 2011 IEEE 11th
International Conference on Data Mining Workshops. 2011;540--547.}

\bibitem[\citeproctext]{ref-subbaswamy2018}
\CSLLeftMargin{14. }%
\CSLRightInline{Subbaswamy A, Schulam P, Saria S.
\href{https://doi.org/10.48550/ARXIV.1812.04597}{Preventing failures due
to dataset shift: Learning predictive models that transport}
{[}Internet{]}. 2018.}

\bibitem[\citeproctext]{ref-subbaswamy2019}
\CSLLeftMargin{15. }%
\CSLRightInline{Subbaswamy A, Saria S.
\href{https://doi.org/10.1093/biostatistics/kxz041}{From development to
deployment: Dataset shift, causality, and shift-stable models in health
AI}. Biostatistics. 2019.}

\bibitem[\citeproctext]{ref-vanAmsterdam2025}
\CSLLeftMargin{16. }%
\CSLRightInline{Amsterdam WAC van. A causal viewpoint on prediction
model performance under changes in case-mix: Discrimination and
calibration respond differently for prognosis and diagnosis predictions
{[}Internet{]}. arXiv; 2024. Available from:
\url{https://arxiv.org/abs/2409.01444}}

\bibitem[\citeproctext]{ref-rothman1976}
\CSLLeftMargin{17. }%
\CSLRightInline{Rothman KJ.
\href{https://doi.org/10.1093/oxfordjournals.aje.a112335}{Causes}. Am J
Epidemiol. 1976;104:587--592.}

\bibitem[\citeproctext]{ref-rothman2005}
\CSLLeftMargin{18. }%
\CSLRightInline{Rothman KJ, Greenland S.
\href{https://doi.org/10.2105/ajph.2004.059204}{Causation and causal
inference in epidemiology}. Am J Public Health. 2005;95:S144--S150.}

\bibitem[\citeproctext]{ref-daniel2024}
\CSLLeftMargin{19. }%
\CSLRightInline{Daniel RM, Farewell DM, Huitfeldt A.
\href{https://doi.org/10.1093/jrsssa/qnae074}{{`}Does god toss logistic
coins?{'} And other questions that motivate regression by composition}.
J R Stat Soc Ser A Stat Soc. 2024;187:636--655.}

\bibitem[\citeproctext]{ref-Flanders2006SCC}
\CSLLeftMargin{20. }%
\CSLRightInline{Flanders WD.
\href{https://doi.org/10.1007/s10654-006-9048-3}{On the relationship of
sufficient component cause models with potential outcome
(counterfactual) models}. Eur J Epidemiol. 2006;21:847--853.}

\bibitem[\citeproctext]{ref-kezios2023}
\CSLLeftMargin{21. }%
\CSLRightInline{Kezios KL, Hayes-Larson E.
\href{https://doi.org/10.3389/fepid.2023.1282809}{Sufficient component
cause simulations: An underutilized epidemiologic teaching tool}. Front
Epidemiol. 2023;3.}

\bibitem[\citeproctext]{ref-Suzuki2011}
\CSLLeftMargin{22. }%
\CSLRightInline{Suzuki E, Yamamoto E, Tsuda T.
\href{https://doi.org/10.1007/s10654-011-9568-3}{Identification of
operating mediation and mechanism in the sufficient-component cause
framework}. Eur J Epidemiol. 2011;26:347--357.}

\bibitem[\citeproctext]{ref-GharbiMeliani2021}
\CSLLeftMargin{23. }%
\CSLRightInline{Gharbi-Meliani A, Dugravot A, Sabia S, et al.
\href{https://doi.org/10.1186/s13195-020-00740-0}{The association of
APOE {\(\varepsilon\)}4 with cognitive function over the adult life
course and incidence of dementia: 20 years follow-up of the whitehall II
study}. Alzheimers Res Ther. 2021;13.}

\bibitem[\citeproctext]{ref-Livingston2024}
\CSLLeftMargin{24. }%
\CSLRightInline{Livingston G, Huntley J, Liu KY, et al.
\href{https://doi.org/10.1016/s0140-6736(24)01296-0}{Dementia
prevention, intervention, and care: 2024 report of the lancet standing
commission}. Lancet. 2024;404:572--628.}

\bibitem[\citeproctext]{ref-Bours2021BayesRule}
\CSLLeftMargin{25. }%
\CSLRightInline{Bours MJL.
\href{https://doi.org/10.1016/j.jclinepi.2020.12.021}{Bayes' rule in
diagnosis}. J Clin Epidemiol. 2021;131:158--160.}

\bibitem[\citeproctext]{ref-Johnson2018BayesRule}
\CSLLeftMargin{26. }%
\CSLRightInline{Johnson KM.
\href{https://doi.org/10.1515/dx-2017-9011}{Erratum to: Using bayes'
rule in diagnostic testing: A graphical explanation}. Diagnosis.
2018;5:89--89.}

\bibitem[\citeproctext]{ref-Altman1994SeSpPrev}
\CSLLeftMargin{27. }%
\CSLRightInline{Altman DG, Bland JM.
\href{https://doi.org/10.1136/bmj.309.6947.102}{Statistics notes:
Diagnostic tests 2: Predictive values}. BMJ. 1994;309:102--102.}

\bibitem[\citeproctext]{ref-Corbo1999ApoeVar}
\CSLLeftMargin{28. }%
\CSLRightInline{Corbo RM, Scacchi R.
\href{https://doi.org/10.1046/j.1469-1809.1999.6340301.x}{Apolipoprotein
{E} (APOE) allele distribution in the world. Is APOE*4 a {``thrifty''}
allele?} Ann Hum Genet. 1999;63:301--310.}

\bibitem[\citeproctext]{ref-Half2009}
\CSLLeftMargin{29. }%
\CSLRightInline{Half E, Bercovich D, Rozen P.
\href{https://doi.org/10.1186/1750-1172-4-22}{Familial adenomatous
polyposis}. Orphanet J Rare Dis. 2009;4.}

\bibitem[\citeproctext]{ref-Lee1999KLDistanceEpi}
\CSLLeftMargin{30. }%
\CSLRightInline{Lee WC.
\href{https://doi.org/10.1093/ije/28.3.521}{Selecting diagnostic tests
for ruling out or ruling in disease: The use of the kullback-leibler
distance}. Int J Epidemiol. 1999;28:521--525.}

\bibitem[\citeproctext]{ref-Joyce2011KLDivergence}
\CSLLeftMargin{31. }%
\CSLRightInline{Joyce JM. Kullback-leibler divergence. In: International
encyclopedia of statistical science {[}Internet{]}. Springer Berlin
Heidelberg; 2011. p. 720--722. Available from:
\url{http://dx.doi.org/10.1007/978-3-642-04898-2_327}}

\bibitem[\citeproctext]{ref-Cover2006}
\CSLLeftMargin{32. }%
\CSLRightInline{Cover TM, Thomas JA. Elements of information theory. 2nd
ed. Nashville, TN: John Wiley \& Sons; 2006.}

\bibitem[\citeproctext]{ref-HippisleyCox2017QRISK3}
\CSLLeftMargin{33. }%
\CSLRightInline{Hippisley-Cox J, Coupland C, Brindle P.
\href{https://doi.org/10.1136/bmj.j2099}{Development and validation of
QRISK3 risk prediction algorithms to estimate future risk of
cardiovascular disease: Prospective cohort study}. BMJ. 2017;j2099.}

\bibitem[\citeproctext]{ref-Chu2009MarkerBivariateMAGLM}
\CSLLeftMargin{34. }%
\CSLRightInline{Chu H, Guo H, Zhou Y.
\href{https://doi.org/10.1177/0272989x09353452}{Bivariate random effects
meta-analysis of diagnostic studies using generalized linear mixed
models}. Med Decis Making. 2009;30:499--508.}

\bibitem[\citeproctext]{ref-Leeflang2012PPVNPVMA}
\CSLLeftMargin{35. }%
\CSLRightInline{Leeflang MMG, Deeks JJ, Rutjes AWS, Reitsma JB, Bossuyt
PMM. \href{https://doi.org/10.1016/j.jclinepi.2012.03.006}{Bivariate
meta-analysis of predictive values of diagnostic tests can be an
alternative to bivariate meta-analysis of sensitivity and specificity}.
J Clin Epidemiol. 2012;65:1088--1097.}

\end{CSLReferences}
\end{document}